\documentclass[aps,prd,twocolumn,a4paper,showpacs,nofootinbib]{revtex4}

\usepackage{amsmath}
\usepackage{amssymb}
\usepackage{graphicx}

\newcommand{\EG}{\textsl{e.g.~}}
\newcommand{\IE}{\textsl{i.e.~}}

\newcommand{\lbulk}{L}

\newcommand{\ads}{$\mathrm{AdS}_{_5}$}
\newcommand{\ue}{\mathrm{e}}
\newcommand{\ud}{\mathrm{d}}
\newcommand{\ub}{\mathrm{b}}

\newcommand{\ui}{\mathrm{i}}

\newcommand{\uR}{\mathrm{R}}
\newcommand{\uT}{\mathrm{T}}
\newcommand{\uY}{\mathrm{Y}}
\newcommand{\uJ}{\mathrm{J}}
\newcommand{\uI}{\mathrm{I}}
\newcommand{\uK}{\mathrm{K}}
\newcommand{\uE}{\mathrm{E}}

\newcommand{\uL}{\mathrm{L}}
\newcommand{\rbr}{r_\ub}
\newcommand{\tbr}{t_\ub}
\newcommand{\sbr}{s_\ub}
\newcommand{\llhh}{\lbulk^2 H^2}
\newcommand{\llhd}{\lbulk^2 \dot{H}}
\newcommand{\kappafive}{\kappa_{_5}}
\newcommand{\kappareduct}{\kappa_{_5}}
\newcommand{\tension}{\mathcal{T}}

\newcommand{\microm}{\,\mathrm{\mu m}}
\newcommand{\mm}{\,\mathrm{mm}}
\newcommand{\planck}{{_\mathrm{P\negthinspace\ell}}}

\newcommand{\bea}{\begin{eqnarray}}
\newcommand{\eea}{\end{eqnarray}}
\newcommand{\be}{\begin{equation}}
\newcommand{\ee}{\end{equation}}
\newcommand{\ra}{\rightarrow}
\newcommand{\Om}{\Omega}

\newcommand{\De}{\Delta}
\newcommand{\sig}{\sigma}
\newcommand{\dd}{\partial}

\newcommand{\Lie}[2]{\mathcal{L}_{#1}{#2}}
\newcommand{\besselj}[1]{j{_{#1}}}
\newcommand{\besselJ}[1]{\uJ{_{#1}}}
\newcommand{\besselY}[1]{\uY{_{#1}}}
\newcommand{\besselI}[1]{\uI{_{#1}}}
\newcommand{\besselK}[1]{\uK{_{#1}}}
\newcommand{\struveL}[1]{\uL{_{#1}}}

\newcommand{\si}[1]{{\scriptscriptstyle{#1}}}

\newcommand{\zero}{{_{0}}}

\newcommand{\four}{{_{4}}}
\newcommand{\five}{{_{5}}}
\newcommand{\ei}{{_{i}}}
\newcommand{\jei}{{_{j}}}
\newcommand{\Esi}{{\si{E}}}

\newcommand{\pertsymbol}[1]{\tilde{#1}}

\newcommand{\spert}{\pertsymbol{s}}
\newcommand{\sbrpert}{\pertsymbol{s}_\ub}

\newcommand{\upert}{\pertsymbol{u}}

\newcommand{\Kpert}{\pertsymbol{K}}
\newcommand{\Xpert}{\pertsymbol{X}}

\newcommand{\qpert}{\pertsymbol{q}}

\newcommand{\Vs}{E}
\newcommand{\Vt}{B}
\newcommand{\Vr}{C}

\newcommand{\Vtgi}{\Xi}
\newcommand{\Vrgi}{\Sigma}

\newcommand{\vs}{e}
\newcommand{\vt}{b}

\newcommand{\vtgi}{\sigma}
\newcommand{\vel}{v}

\newcommand{\velgi}{\vartheta}
\newcommand{\VtgiT}{\Vtgi_{\si{\uT}}}
\newcommand{\VtgiR}{\Vtgi_{\si{\uR}}}
\newcommand{\VrgiT}{\Vrgi_{\si{\uT}}}
\newcommand{\VrgiR}{\Vrgi_{\si{\uR}}}

\newcommand{\aniso}{\pi}
\newcommand{\locpert}{\Upsilon}

\newcommand{\emitime}{{_{\mathrm{E}}}}
\newcommand{\Ho}{H_\zero}

\newcommand{\az}{a_\zero}
\newcommand{\etao}{\eta_\zero}
\newcommand{\etae}{\eta_\emitime}
\newcommand{\vv}{\overline{\eta}}
\newcommand{\vo}{\vv_\zero}
\newcommand{\ve}{\vv_\emitime}
\newcommand{\kb}{\bar{k}}
\newcommand{\Ao}{A_\zero}
\newcommand{\kbe}{\kb_\emitime}
\newcommand{\kbl}{\kb_\ell}
\newcommand{\xx}{x}
\newcommand{\xe}{\xx_\emitime}
\newcommand{\ze}{z_\emitime}
\newcommand{\Cl}{C_\ell}
\newcommand{\Ic}{\mathcal{I}}

\newcommand{\yl}{y_\ell}

\begin{document}

\title{CMB anisotropies from vector perturbations in the bulk}

\author{Christophe Ringeval}
\email{christophe.ringeval@physics.unige.ch}
\affiliation{D\'epartement de Physique Th\'eorique, Universit\'e de
Gen\`eve, 24 quai Ernest Ansermet, 1211 Gen\`eve 4, Switzerland.}

\author{Timon Boehm}
\email{timon.boehm@physics.unige.ch}
\affiliation{D\'epartement de Physique Th\'eorique, Universit\'e de
Gen\`eve, 24 quai Ernest Ansermet, 1211 Gen\`eve 4, Switzerland.}

\author{Ruth Durrer}
\email{ruth.durrer@physics.unige.ch}
\affiliation{D\'epartement de Physique Th\'eorique, Universit\'e de
Gen\`eve, 24 quai Ernest Ansermet, 1211 Gen\`eve 4, Switzerland.}

\date{July 10, 2003}

\begin{abstract}
The vector perturbations induced on the brane by gravitational waves
propagating in the bulk are studied in a cosmological
framework. Cosmic expansion arises from the brane motion in a
non-compact five-dimensional anti-de Sitter spacetime. By solving the
vector perturbation equations in the bulk, for generic initial
conditions, we find that they give rise to growing modes on the brane
in the Friedmann-Lema\^{\i}tre era. Among these modes, we exhibit a
class of normalizable perturbations, which are exponentially growing
with respect to conformal time on the brane.  The presence of these
modes is, at least, strongly constrained by the current observations
of the cosmic microwave background (CMB). We estimate the anisotropies
they induce in the CMB, and derive quantitative constraints on the
allowed amplitude of their primordial spectrum. Our results provide 
stringent constraints for all braneworld models with bulk inflation.

\end{abstract}

\maketitle

\section{Introduction}

The idea that our universe may have more than three spatial dimensions
has been originally introduced by Nordstr\"om~\cite{Nordstrom:1914},
Kaluza~\cite{Kaluza:1921} and Klein~\cite{Klein:1926}. The fact that
super string theory, the most promising candidate for a theory of
quantum gravity, is consistent only in ten spacetime dimensions (11
dimensions for M-theory) has led to a revival of these
ideas~\cite{Polchinski:1998rq,Polchinski:1998rr,Horava:1996qa}. It has
also been found that string theories naturally predict lower
dimensional ``branes" to which fermions and gauge particles are
confined, while gravitons (and the dilaton) propagate in the
bulk~\cite{Antoniadis:1990ew,Polchinski:1995mt,Lukas:1998qs}. Such
``braneworlds" have been studied in a phenomenological way already before the
discovery that they are actually realized in string
theory~\cite{Akama:1982jy,Rubakov:1983bb}.

Recently it has been emphasized that relatively large extra-dimensions
(with typical length $\lbulk \simeq \microm$) can ``solve" the
hierarchy problem: The effective four dimensional Newton constant
given by $G_\four \propto G/\lbulk^N$ can become very small even if
the fundamental gravitational constant $ G \simeq
m_{\planck}^{-(2+N)}$ is of the order of the electro-weak scale. Here
$N$ denotes the number of
extra-dimensions~\cite{Arkani-Hamed:1998rs,Arkani-Hamed:1998nn,Antoniadis:1998ig,Randall:1999ee}.
It has also been shown that extra-dimensions may even be infinite
if the geometry contains a so-called ``warp
factor"~\cite{Randall:1999vf}.

The size of the extra-dimensions is constrained by the requirement of
recovering usual four-dimensional Einstein gravity on the brane, at
least on scales tested by
experiments~\cite{Long:2002wn,Uzan:2000mz,Allen:2000ih}.  Models with
either a small Planck mass in the
bulk~\cite{Arkani-Hamed:1998rs,Arkani-Hamed:1998nn,Antoniadis:1998ig},
or with non-compact warped
extra-dimensions~\cite{Randall:1999ee,Randall:1999vf}, have until now
been shown to lead to an acceptable cosmological phenomenology on the
brane~\cite{Binetruy:1999ut,Csaki:1999jh,Cline:1999ts,Shiromizu:1999wj,
Flanagan:1999cu,Maartens:1999hf,Rubakov:2001kp}, with or without $Z_2$
symmetry in the bulk~\cite{Carter:2001nj,Battye:2001yn,Carter:2001af}.
Explicite cosmological scenarii leading to a nearly
Friedman-Lema\^{\i}tre universe at late time have been proposed by
means of a static brane in a time dependent
bulk~\cite{Binetruy:1999hy,Vollick:1999uz} or with a moving brane in
an anti-de Sitter bulk~\cite{Kraus:1999it,Ida:1999ui}. It has been
shown that both approaches are actually
equivalent~\cite{Mukohyama:1999wi}.

One can also describe braneworlds as topological defects in the
bulk~\cite{Bonjour:1999kz,Gregory:1999gv,Antunes:2002hn,Nihei:2000gb,
Gherghetta:2000qi}. This approach is equivalent to the geometrical one
in the gravity sector~\cite{Ringeval:2001cq}, while it admits an
explicite mechanism to confine matter and gauge fields on the
brane~\cite{Ringeval:2001cq,Ringeval:2001xd,Bajc:1999mh,Dvali:1997xe,
Dubovsky:2000am,Dvali:2000rx,Dimopoulos:2000ej,Duff:2000jk,Oda:2001ux,
Ghoroku:2001zu,Akhmedov:2001ny,Neronov:2001qv}. Although depending on
the underlying theory, the stability studies of these defects have
shown that dynamical instabilities appear on the brane when there are
more than one non-compact
extra-dimensions~\cite{Gherghetta:2000jf,Kanti:2001vb,Peter:2003zg},
whereas this is not the case for a five-dimensional
bulk~\cite{Giovannini:2001fh} provided that the usual fine-tuning
between model parameters is fixed~\cite{Boehm:2001sp}.

The next step is now to derive observational consequences of
braneworld cosmological models, \EG the anisotropies of the cosmic
microwave background (CMB). To that end, a lot of work has recently
been invested to derive gauge invariant perturbation theory in
braneworlds with one
co-dimension~\cite{Sasaki:1999mi,Langlois:2000ia,vandeBruck:2000ju,Bridgman:2001mc}. According
to the background cosmology, the perturbations can be derived for a
brane at rest~\cite{Riazuelo:2002mi}, or moving in a perturbed anti-de
Sitter
spacetime~\cite{Mukohyama:1999wi,Mukohyama:2000ga,Mukohyama:2000ui,
Mukohyama:2001yp,Deruelle:2000yj,Deffayet:2002fn}. Whatever the
approach chosen, the perturbation equations are quite cumbersome and
it is difficult to extract interesting physical consequences
analytically. Also the numerical treatment is much harder than in
usual four dimensional perturbation theory, since it involves partial
differential equations.

Nevertheless, it is useful to derive some simple physical consequences
of perturbation theory for braneworlds before performing intensive
numerical studies. This has been done for tensor perturbations on
the brane in a very phenomenological way in Ref.~\cite{Leong:2002hs}
or on a more fundamental level in Ref.~\cite{Frolov:2002va}. 
Tensor modes in the bulk which induce scalar perturbations on the
brane have been studied in Ref.~\cite{Durrer:2003rg} and let to important
constraints for braneworlds.

In this article we consider a braneworld in a five dimensional bulk
where cosmology is induced by the motion of a ``3-brane'' in \ads. The
bulk perturbation equations are considered without bulk sources and
describe gravity waves in the bulk. The present work concentrates on
the part of these gravity waves which results in vector perturbations
on the brane.

For the sake of clarity, we first recall how cosmology on the brane
can be obtained via the junction conditions, particularly emphasizing
how $Z_2$ symmetry is
implemented~\cite{Binetruy:1999ut,Csaki:1999jh,Cline:1999ts,Shiromizu:1999wj,
Flanagan:1999cu,Maartens:1999hf,Rubakov:2001kp}. After re-deriving the
bulk perturbation equations for the vector components in terms of
gauge invariant
variables~\cite{Mukohyama:1999wi,Mukohyama:2000ga,Mukohyama:2000ui,
Mukohyama:2001yp,Deruelle:2000yj}, we analytically find the most
general solutions for arbitrary initial conditions. The time evolution
of the induced vector perturbations on the brane is then derived by
means of the perturbed junction conditions. The main result of the
paper is that vector perturbations in the bulk generically give rise
to vector perturbations on the brane which grow either as a power law
or even exponentially with respect to conformal time. This behavior
differs essentially from the usual decay of vector modes obtained in
standard four-dimensional cosmology, and may lead, at least, to
observable effects of extra-dimensions in the CMB.

In the next section, the cosmological braneworld model obtained by the
moving brane in an anti-de Sitter bulk is briefly recalled. In
Sect.~\ref{sec:bulkpert} we set up the vector perturbation equations
and solve them in the bulk. In Sect.~\ref{sec:branepert} the induced
perturbations on the brane are derived and compared to the usual
four-dimensional ones, while Sect.~\ref{sec:cmb} deals with the
consequences of these new results on CMB anisotropies. The resulting
new constraints for viable braneworlds are discussed in the
conclusion.


\section{Background}

As mentioned in the introduction, our universe is considered to be a
3-brane embedded in five dimensional anti-de Sitter spacetime with
flat three dimensional spatial sections
\begin{equation}
\label{eq:metric} \ud s^2 = g_{\si{AB}} \ud x^{\si{A}} \ud
x^{\si{B}}=\dfrac{r^2}{\lbulk^2} \left(-\ud t^2 + \delta_{ij} \ud
x^i \ud x^j \right) + \dfrac{\lbulk^2}{r^2} \ud r^2.
\end{equation}
The capital Latin indices $A,B$ run from $0$ to $4$ and $i,j$ from
$1$ to $3$. Anti-de Sitter spacetime is a solution of Einstein's
equations with a negative cosmological constant $\Lambda$
\begin{equation}
\label{eq:einstein} G_{\si{AB}} + \Lambda g_{\si{AB}} = 0,
\end{equation}
provided that the curvature radius $\lbulk$ satisfies
\begin{equation}
\label{eq:Lambda}  \lbulk^2 = -\dfrac{6}{\Lambda}.
\end{equation}

Another coordinate system for anti-de Sitter space which is often used
in braneworld models is defined by $r^2/\lbulk^2 = e^{-2 \varrho
/\lbulk}$ so that \be \label{eq:adsrs} \ud s^2 = g_{\si{AB}} \ud
x^{\si{A}} \ud x^{\si{B}}=e^{-2 \varrho /\lbulk} \left(-\ud t^2 + \delta_{ij}
\ud x^i \ud x^j \right) + \ud \varrho^2.  \ee


\subsection{Embedding and motion of the brane}

The position of the brane in the \ads bulk is given by
\begin{equation}
x^{\si{M}}=X^{\si{M}}(y^{\mu}),
\end{equation}
where $X^{\si{M}}$ are embedding functions depending on the internal
brane coordinates $y^\mu$ ($\mu =0, \cdots, 3)$. Making use of
reparametrization invariance on the brane, we choose $x^i = X^i =
y^i$. The other embedding functions are written
\begin{equation}
\label{eq:embedding}
\begin{aligned}
 X^0 = \tbr(\tau), && X^4 = \rbr(\tau),
\end{aligned}
\end{equation}
where $\tau \equiv y^0$ denotes cosmic time on the brane. Since we
want to describe a homogeneous and isotropic brane, $X^0$ as well as
$X^4$ are required to be independent of the spatial coordinates $y^i$.
The four tangent vectors to the brane are given by
\begin{equation}
\label{eq:tangent}
e^{\si{M}}_\mu\dd_{\si{M}}=\dfrac{\partial X^{\si{M}}}{\partial
y^\mu}\dd_{\si{M}},
\end{equation}
and the unit space-like normal 1-form $n_{\si{M}}$ is defined (up to a
sign) by the orthogonality and normalization conditions
\begin{equation}
\label{eq:normal}
\begin{aligned}
n_{\si{M}} e^{\si{M}}_\mu &= 0, & g^{\si{AB}} n_{\si{A}} n_{\si{B}}&=1.
\end{aligned}
\end{equation}
Adopting the sign convention so that $n$ points in the direction in which
the brane is moving (growing $\rbr$ for an expanding universe), one
finds using
\begin{equation}
e^0_\tau  =\dot{\tbr},\quad e^4_\tau =\dot{\rbr}, \quad e^i_j =\delta^i_j,
\end{equation}
the components of the normal
\begin{equation}
\label{eq:vectors}
n_0 = -\dot{\rbr}, \quad n_4 = \dot{\tbr}, \quad n_i =0.
\end{equation}
The other components are vanishing, and the dot denotes
differentiation with respect to the brane time $\tau$.

This embedding ensures that the induced metric on the brane describes
a spatially flat homogeneous and isotropic universe,
\begin{equation}
\label{eq:inducedmetric}
\ud \sbr^2 = q_{\mu \nu} \ud y^{\mu} \ud y^{\nu} = - \ud \tau^2 +
a^2(\tau) \delta_{ij} \ud y^i \ud y^j,
\end{equation}
where $a(\tau)$ is the usual scale factor, and $q_{\mu \nu}$ is
the pull-back of the bulk metric onto the brane
\begin{equation}
\label{eq:qmunu} q_{\mu \nu} = g_{\si{AB}} e^{\si{A}}_\mu
e^{\si{B}}_\nu,
\end{equation}
(see \EG \cite{Carter:1997pb,Ringeval:2002qi}).
The first fundamental form $q_{\si{AB}}$ is now defined by
\begin{equation}
\label{eq:firstfund}
q^{\si{AB}}=q^{\mu \nu} e^{\si{A}}_\mu e^{\si{B}}_\nu,
\end{equation}
\IE the push-forward of the inverse of the induced metric tensor
\cite{Carter:1999hx,Carter:1997pb}. One can also define an orthogonal
projector onto the brane which can be expressed in terms of the normal
1-form
\begin{equation}
\label{eq:orthoproj} \perp_{\si{AB}}=n_{\si{A}}
n_{\si{B}}=g_{\si{AB}} - q_{\si{AB}},
\end{equation}
in the case of only one codimension.

Upon inserting the equations (\ref{eq:metric}), (\ref{eq:vectors}) and
(\ref{eq:firstfund}) into the above equation, one finds a parametric
form for the brane 
trajectory~\cite{Kraus:1999it,Ida:1999ui,Mukohyama:2001yp,Deruelle:2000yj}
\begin{equation}
\label{eq:branemotion}
\begin{aligned}
\rbr(\tau) &= a(\tau) \lbulk, \\
\dot{\tbr}(\tau) &= \dfrac{1}{a} \sqrt{1 + \llhh},
\end{aligned}
\end{equation}
where $H=\dot{a}/a$ denotes the Hubble parameter on the brane.
Alternatively, comparing expression (\ref{eq:qmunu}) with the
Friedmann metric~(\ref{eq:inducedmetric}) yields the same result.

Therefore, the unperturbed motion induces a cosmological
expansion on the 3-brane if $\rbr$ is growing with $\tbr$.


\subsection{Extrinsic curvature and unperturbed junction conditions}
\label{ssect:backjunct}

The Lanczos--Sen--Darmois--Israel junction conditions\footnote{In the
following, they will be simply referred to as ``junction
conditions''.} govern the evolution of the brane. They relate the jump
of the extrinsic curvature across the brane to its surface energy-momentum
content~\cite{Lanczos:1924,Sen:1924,Darmois:1927,Israel:1966}. The
extrinsic curvature tensor projected on the brane can be expressed by
\begin{equation}
\label{eq:extcurv} K_{\mu \nu} = - e^{\si{A}}_\mu
e^{\si{B}}_\nu \nabla_{\si{A}} n_{\si{B}} =
-\dfrac{1}{2}e^{\si{A}}_\mu e^{\si{B}}_\nu \Lie{n}{g_{\si{AB}}}.
\end{equation}
Here $\nabla$ denotes the covariant derivative with respect to the
bulk metric, and $\Lie{n}{}$ is the five dimensional Lie-derivative in
the direction of the unit normal on the brane. With the sign choice in
Eq.~(\ref{eq:extcurv}), the junction conditions
read~\cite{Misner:1970aa}
\begin{equation}
\label{eq:jump} K_{\mu \nu}^{>} - K_{\mu \nu}^{<} = \kappareduct^2
\left( S_{\mu \nu} - \dfrac{1}{3} S q_{\mu \nu} \right) \equiv
\kappareduct^2 \widehat{S}_{\mu \nu},
\end{equation}
where $S_{\mu \nu}$ is the energy momentum tensor on the brane with
trace $S$, and
\begin{equation}
\kappafive^2 \equiv 6\pi^2 G_\five=\dfrac{1}{M_\five^{3}}, 
\end{equation}
with $M_\five$ and $G_\five$, the five dimensional (fundamental)
Planck mass and Newton constant. The superscripts ``$>$'' and ``$<$''
stand for the bulk sides with $r > \rbr$ and $r < \rbr$. As already
noticed, the brane normal vector $n^{\si{M}}$ points into the
direction of increasing $r$ [see
Eq.~(\ref{eq:vectors})]. Eq.~(\ref{eq:jump}) is usually referred to as
second junction condition. The first junction condition simply states
that the first fundamental form (\ref{eq:firstfund}) is continuous
across the brane.

In general, there is a force acting on the brane which is due to its
curvature in the higher dimensional geometry. It is given by the
contraction of the brane energy momentum tensor with the average
of the extrinsic curvature on both sides of the brane~\cite{Battye:2001yn}
\begin{equation}
\label{eq:mean} S^{\mu \nu} \left(K_{\mu \nu}^{>} + K_{\mu
\nu}^{<} \right) = 2 f = 0.
\end{equation}
This force $f$, normal to the brane, is exerted by the asymmetry of
the bulk with respect to the
brane~\cite{Carter:1997pb,Battye:2001yn}. In this paper, we consider
only the case in which the bulk is $Z_2$ symmetric across the brane,
so that $f=0$. In this case the motion of the brane is caused 
by the stress energy tensor of the brane itself which is exactly the
cosmological situation we have in mind.

From Eqs.~(\ref{eq:vectors}), (\ref{eq:inducedmetric}),
(\ref{eq:branemotion}) and (\ref{eq:extcurv}), noting that the
extrinsic curvature can be expressed purely in terms of the 
internal brane coordinates~\cite{Deruelle:2000yj,Mukohyama:2001yp},
one has
\begin{equation}
\label{eq:extusefull}
\begin{aligned}
K_{\mu \nu} &= -\dfrac{1}{2} \left[g_{\si{AB}}\left(e^{\si{A}}_\mu
\partial_\nu n^{\si{B}} + e^{\si{A}}_\nu  \partial_\mu
n^{\si{B}}\right) + e^{\si{A}}_\mu e^{\si{B}}_\nu n^{\si{C}}
g_{\si{AB},\si{C}} \right]~.
\end{aligned}
\end{equation}
A short computation shows that the  non-vanishing components of the
extrinsic curvature are
\begin{equation}
\label{eq:kcomponents}
\begin{aligned}
K_{\tau \tau} &= \dfrac{1+ \llhh + \llhd}{\lbulk \sqrt{1 + \llhh}},\\
K_{ij} &= - \dfrac{a^2}{\lbulk}\sqrt{1+\llhh} \delta_{ij}.
\end{aligned}
\end{equation}
It is clear, that the extrinsic curvature evaluated at some brane
position $\rbr$ does not jump if the presence of the brane does not
modify anti-de Sitter space. Like in the Randall--Sundrum
model~\cite{Randall:1999vf}, in order to accommodate cosmology, the
bulk space-time structure is modified by gluing the mirror symmetric
of anti-de Sitter space on one side of the brane onto the
other~\cite{Mukohyama:1999wi}. There are two possibilities: one can
keep the ``$r>\rbr$'' side and replace the ``$r<\rbr$'' side to get
\begin{equation}
\begin{aligned}
K_{\mu \nu}^{>}&=K_{\mu \nu}, & K_{\mu \nu}^{<} &=-K_{\mu \nu},
\end{aligned}
\end{equation}
where $K_{\mu \nu}$ is given by
Eq.~(\ref{eq:kcomponents}). Conversely, keeping the $r<\rbr$ side
leads to
\begin{equation}
\begin{aligned}
K_{\mu \nu}^{>}&=-K_{\mu \nu}, & K_{\mu \nu}^{<} &=K_{\mu \nu}.
\end{aligned}
\end{equation}
Note that both cases verify the force equation (\ref{eq:mean}). From
the time and space components of the junction conditions
(\ref{eq:jump}) one obtains, respectively
\begin{align}
\label{eq:friedmanT}
\pm \dfrac{1+ \llhh + \llhd}{\lbulk \sqrt{1 + \llhh}} &= \dfrac{1}{2}
\kappareduct^2 \left( P + \rho\right) - \dfrac{1}{6} \kappareduct^2
\left(\rho + \tension \right),\\
\label{eq:friedmanS}
\pm \dfrac{\sqrt{1+\llhh}}{\lbulk} &= -\dfrac{1}{6} \kappareduct^2
\left(\rho + \tension\right).
\end{align}
Here the brane stress tensor is assumed to be that of a cosmological
fluid plus a pure tension $\tension$, \IE
\begin{equation}
\label{eq:branetensor} S_{\mu \nu} = \left(P + \rho \right) u_\mu
u_\nu + P q_{\mu \nu} - \tension q_{\mu \nu},
\end{equation}
$\rho$ and $P$ being the usual energy density and pressure on the
brane, and $u^\mu$ the comoving four-velocity. The ``$\pm$'' signs in
Eqs.~(\ref{eq:friedmanT}) and (\ref{eq:friedmanS}) are obtained by
keeping, respectively, the $r>\rbr$, or $r<\rbr$, side of the bulk. In
order to allow for a positive total brane energy density, $\rho+
\tension$, we have to keep the $r<\rbr$ side and glue it symmetrically
on the $r>\rbr$ one\footnote{Note that we obtain the same result as in
Ref.~\cite{Deruelle:2000yj}, \IE a positive brane tension for an
expanding universe is obtained by keeping the anti-de Sitter side
which is behind an expanding brane, \emph{according to its motion}.}.
In the trivial non-expanding case this construction reproduces the
Randall-Sundrum solution with warp factor $\exp(-|\varrho|/\lbulk)$,
for $-\infty<\varrho<\infty$. In our coordinates, we just have
$0<r\le \rbr$ on either side of the brane, and the bulk is now
described by two copies of the ``bulk behind the brane". Even if $r$
only takes values inside a finite interval, and even though the volume
of the extra dimension,
\begin{equation}
V=2\int_0^{\rbr}\sqrt{|g|} \, \ud r = \dfrac{\rbr}{2}
\left(\dfrac{\rbr}{\lbulk}\right)^3,
\end{equation}
 is finite, the bulk is non-compact and its spectrum of perturbation
modes has no gap (like in the Randall-Sundrum model).

From Eqs.~(\ref{eq:friedmanT}) and (\ref{eq:friedmanS}), one can check
that energy conservation on the brane is verified
\begin{equation}
\label{eq:enercons}
\dot{\rho} + 3H \left(P + \rho\right) = 0.
\end{equation}
Solving Eq.~(\ref{eq:friedmanS}) for the Hubble parameter yields
\begin{equation}
\label{eq:hubbleevol}
H^2 = \dfrac{\kappareduct^4 \tension}{18} \rho \left(1 + \dfrac{\rho}{2
\tension} \right) + \dfrac{\kappareduct^4}{36} \tension^2 -
\dfrac{1}{\lbulk^2}.
\end{equation}
At ``low energies", $|\rho/\tension| \ll 1$, the usual Friedmann
equation is recovered provided the fine-tuning condition
\begin{equation}
\dfrac{\kappareduct^4}{36} \tension^2 = \dfrac{1}{\lbulk^2},
\end{equation}
is satisfied. The four dimensional Newton constant is then given by
\begin{equation}
\kappa_\four^2 \equiv 8\pi G_\four =\dfrac{\kappareduct^4
\tension}{6},
\end{equation}
requiring a positive tension in order to get a positive effective
four-dimensional Newton constant. Note also that low energy simply means
$\tau^2 \sim H^{-2} \gg L^2$. In the Friedmann-Lema\^{\i}tre era,
the solution of Eq.~(\ref{eq:hubbleevol}) reads
\begin{equation}
\label{eq:nowhubbleevol}
\begin{aligned}
H & \simeq \Ho \left(\dfrac{a}{\az}\right)^{-3(1+w)/2},\\ \dot{H}
& \simeq -\dfrac{3}{2} (1+w) \Ho^2 \left(\dfrac{a}{\az}\right)^{-3(1+w)},
\end{aligned}
\end{equation}
for a cosmological equation of state $P=w \rho$ with constant
$w$. The parameters $H_\zero$ and $\az$ refer, respectively, to the
Hubble parameter and the scale factor today. For the matter era we have
$w=0$, and during the radiation era $w=1/3$.


\section{Gauge invariant perturbation equations in the bulk}
\label{sec:bulkpert}

A general perturbation in the bulk can be decomposed into
``3-scalar'', ``3-vector'' and ``3-tensor'' parts which are
irreducible components under the group of isometries (of the
unperturbed space time) $SO(3)\times E_3$, the group of three
dimensional rotations and translations. In this paper we restrict
ourselves to 3-vector perturbations\footnote{For sake of clarity, the
prefix ``3-'' will be dropped in what follows, and the term ``vector"
will be always applied here for spin 1 with respect to the surfaces of
constant $t$ and $r$.} and consider an ``empty bulk", i.e. the case
where there are no sources in the bulk except for a negative
cosmological constant. Therefore we consider only bulk gravity waves
since they are the only modes present when the energy momentum tensor
is not perturbed. It is well known (see \EG
Ref.~\cite{Riazuelo:2002mi}) that gravity waves in $4+1$ dimensions
have five degrees of freedom which can be decomposed with respect to
their spin in $3+1$ dimensions into a spin $2$ field, the ordinary
graviton, a spin $1$ field, often called the gravi-photon and into a
spin $0$ field, the gravi-scalar. In this work we study the evolution
of the gravi-photon in the background described in the previous
section.

After setting up our notations, we define gauge invariant vector
perturbation variables in the bulk and write down the corresponding
Einstein equations. Analytic solutions for all vector modes in the
bulk are then derived.


\subsection{Bulk perturbation variables}

Considering only vector perturbations in the bulk, the five
dimensional perturbed metric can be parameterized as
\begin{equation}
\label{eq:bulkpertmetric}
\begin{aligned}
\ud \spert^2 & = -\dfrac{r^2}{\lbulk^2} \ud t^2 +
\dfrac{r^2}{\lbulk^2} \left(\delta_{ij} + \nabla_i \Vs_j + \nabla_j
\Vs_i \right) \ud x^i \ud x^j \\ & + \dfrac{\lbulk^2}{r^2}\ud r^2 +
2 \Vt_i \dfrac{r^2}{\lbulk^2} \ud t \ud x^i + 2 \Vr_i \ud x^i \ud r,
\end{aligned}
\end{equation}
where $\nabla_i$ denotes the connection in the three dimensional
subspace of constant $t$ and constant $r$. Assuming this space to be 
flat we simply have $\nabla_i = \partial_i$. The quantities 
$\Vs^i$, $\Vt^i$, and $\Vr^i$ are
divergenceless vectors \IE $\partial_i \Vs^i = \partial_i
\Vt^i =\partial_i\Vr^i=  0$.

As long as we want to solve for the vector perturbations in the bulk
only, the presence of the brane is not yet relevant. Later it will
appear as a boundary condition for the bulk perturbations via the
junction conditions that will be discussed in
Sect.~\ref{subsec:perturbedjunction}.

Under a linearized vector type coordinate transformation in the bulk, 
$x^{\si{M}} \rightarrow x^{\si{M}} + \varepsilon^{\si{M}}$, with
$\varepsilon_{\si{M}}=(0,\varepsilon_i,0)$, the perturbation variables
defined above transform as
\begin{equation}
\label{eq:bulktrans}
\begin{aligned}
\Vs_i &\rightarrow  \Vs_i + \dfrac{\lbulk^2}{r^2} \varepsilon_i,\\
\Vt_i &\rightarrow  \Vt_i + \dfrac{\lbulk^2}{r^2} \partial_t \varepsilon_i,\\
\Vr_i &\rightarrow  \Vr_i + \partial_r \varepsilon_i -
\dfrac{2}{r} \varepsilon_i.
\end{aligned}
\end{equation}

As expected for three divergenceless vector variables and one
divergenceless vector type gauge transformation, there remain four
degrees of freedom which can be chosen as the two gauge invariant
vectors
\begin{align}
\label{eq:defVtgi}
\Vtgi_i & = \Vt_i - \partial_t \Vs_i,\\
\label{eq:defVrgi}
\Vrgi_i & = \Vr_i - \dfrac{r^2}{\lbulk^2} \partial_r \Vs_i.
\end{align}
Note that in the gauge $\Vs_i=0$ these gauge-invariant variables
simply become $ \Vt_i$ and $\Vr_i$ respectively.


\subsection{Bulk perturbation equations and solutions}

A somewhat cumbersome derivation of the Einstein tensor from
(\ref{eq:bulkpertmetric}) to first order in the perturbations leads
via  Eqs.~(\ref{eq:einstein}) to the following vector perturbation equations,
\begin{align}
\label{eq:bulkpertconst}
\partial_t \Vtgi - \dfrac{\lbulk}{r} \partial_r \negthickspace
\left(\dfrac{r^3}{\lbulk^3} \Vrgi \right)  = 0,\\
\label{eq:bulkpertVt}
\dfrac{r^4}{\lbulk^2} \partial_r^2 \Vtgi + 5 \dfrac{r^3}{\lbulk^2}
\partial_r \Vtgi  - \lbulk^2 \partial_t^2 \Vtgi + \lbulk^2 \Delta \Vtgi  =
0, \\
\label{eq:bulkpertVr}
\dfrac{r^4}{\lbulk^2} \partial_r^2 \negthickspace
\left(\dfrac{r^3}{\lbulk^3} \Vrgi \right) - \dfrac{r^3}{\lbulk^2}
\partial_r \negthickspace \left(\dfrac{r^3}{\lbulk^3} \Vrgi \right) -
\lbulk^2 \partial_t^2 \negthickspace \left(\dfrac{r^3}{\lbulk^3} \Vrgi
\right) \nonumber \\ + \lbulk^2 \Delta \negthickspace
\left(\dfrac{r^3}{\lbulk^3} \Vrgi \right) = 0,
\end{align}
where $\Delta$ denotes the spatial Laplacian, \IE
\begin{equation}
\Delta = \delta^{ij} \partial_i \partial_j,
\end{equation}
and the spatial index on the perturbations has been omitted. One can
check that these equations are consistent, \EG with the master
function approach of Ref.~\cite{Mukohyama:2000ui}.

A complete set of solutions for these equations can easily be found by
Fourier transforming with respect to $x^i$, and making the separation
ansatz:
\begin{eqnarray}
\label{eq:separation}
\Vtgi(t,r,{\mathbf{k}}) &=& \VtgiT(t,{\mathbf{k}})
\VtgiR(r,{\mathbf{k}}), \\ \Vrgi(t,r,{\mathbf{k}}) &=&
\VrgiT(t,{\mathbf{k}}) \VrgiR(r,{\mathbf{k}}).
\end{eqnarray}
The most general solution is simply a linear combination of such
elementary modes. Eq.~(\ref{eq:bulkpertVt}) now
leads to two ordinary differential equations for $\VtgiT$ and
$\VtgiR$,
\begin{align}
\label{eq:VtgiR}
{r^4} \dfrac{\partial_r^2 \VtgiR}{\VtgiR} + 5
{r^3} \dfrac{\partial_r \VtgiR}{\VtgiR} & = \pm \lbulk^2\Om^2,\\
\label{eq:VtgiT}
\dfrac{\partial_t^2 \VtgiT}{\VtgiT} + k^2 & = \pm \Om^2,
\end{align}
where $k$ is the spatial wave number, and $\pm\Om^2$ the separation
constant having the dimension of an inverse length squared. We choose
$\Om^2\ge 0$ and indicate the two signs in Eqs.~(\ref{eq:VtgiR}) and
(\ref{eq:VtgiT}) by ``$\pm$''. The frequency $\Om$ represents the rate
of change of $\VtgiR$ at $r\sim \lbulk$, while the rate of change of
$\VtgiT$ is $\sqrt{| \Om^2 \mp k^2|}$.  Eq.~(\ref{eq:VtgiR}) is a
Bessel differential equation of order two for the ``$-$'' sign and
a modified Bessel equation of order two for the ``$+$''
sign~\cite{Abramovitz:1970aa}, while Eq.~(\ref{eq:VtgiT}) exhibits,
respectively, oscillatory or exponential evolution with respect to the
bulk time. From Eq.~(\ref{eq:bulkpertVr}), similar equations are
derived for $\VrgiT(t,\bf{k})$ and $ \VrgiR(r,\bf{k})$.  This time,
the radial function is given by Bessel functions of order one. The
constraint equation (\ref{eq:bulkpertconst}) ensures that the
separation constant $\pm \Om^2$ is the same for both vectors and it
also determines the proportionality factor of the amplitudes. The
general solution of Eqs.~(\ref{eq:bulkpertconst}) to
(\ref{eq:bulkpertVr}) is a superposition of modes $(\Om, {\bf k})$
which are given by
\begin{align}
\label{eq:bulksolVt}
\Vtgi &\propto \left\{
\begin{aligned}
\dfrac{\lbulk^2}{r^2} &\besselK{2} \negthickspace \left(\dfrac{\lbulk^2}{r}\Om \right)
\ue^{\displaystyle \pm t\sqrt{\Om^2-k^2}}  \\
\dfrac{\lbulk^2}{r^2} &\besselI{2}\negthickspace\left(\dfrac{\lbulk^2}{r}\Om\right)
\ue^{\displaystyle \pm t \sqrt{\Om^2-k^2}} \\
\dfrac{\lbulk^2}{r^2} &\besselJ{2}\negthickspace\left(\dfrac{\lbulk^2}{r}\Om \right)
\ue^{\displaystyle \pm i t\sqrt{\Om^2+k^2}} \\
\dfrac{\lbulk^2}{r^2} &\besselY{2}\negthickspace\left(\dfrac{\lbulk^2}{r}\Om \right)
\ue^{\displaystyle \pm i t\sqrt{\Om^2+k^2}} 
\end{aligned} \right.\\
\label{eq:bulksolVr}
\Vrgi &\propto \left\{
\begin{aligned}
\mp \sqrt{1-k^2/\Om^2} &\dfrac{\lbulk^2}{r^2}  \besselK{1}\negthickspace
\left(\dfrac{\lbulk^2}{r}\Om\right)
\ue^{\displaystyle \pm t\sqrt{\Om^2-k^2}}  \\
\mp \sqrt{1-k^2/\Om^2} &\dfrac{\lbulk^2}{r^2}  \besselI{1}\negthickspace
\left(\dfrac{\lbulk^2}{r}\Om\right)
\ue^{\displaystyle \pm t\sqrt{\Om^2-k^2}}  \\
\pm \sqrt{1+k^2/\Om^2} &\dfrac{\lbulk^2}{r^2}  \besselJ{1}\negthickspace
\left(\dfrac{\lbulk^2}{r} \Om \right)
i \ue^{\displaystyle \pm i t \sqrt{\Om^2+k^2}} \\
\pm \sqrt{1+k^2/\Om^2} &\dfrac{\lbulk^2}{r^2}   \besselY{1}\negthickspace
\left(\dfrac{\lbulk^2}{r}\Om\right)
i \ue^{\displaystyle \pm i t \sqrt{\Om^2+k^2}} 
\end{aligned} \right.
\end{align}

Here $\besselK{p}$ and $\besselI{p}$ are the modified Bessel functions
of order $p$ while $\besselJ{p}$ and $\besselY{p}$ are the ordinary
ones.  The $\pm$ signs in
Eqs.~(\ref{eq:bulksolVt}) and (\ref{eq:bulksolVr}) come from the two
independent solutions stemming from the second order differential
equation (\ref{eq:VtgiT}), which is independent of the sign of the
separation constant. In general, each of these 
modes\footnote{In the following, they
will be labelled by the kind of Bessel function they involve, \EG
``K--mode'', ``I--mode'' \dots} can be multiplied by a proportionality
coefficient which depends on the wave vector $\mathbf{k}$ and the
separation constant $\pm\Om^2$ and which has to be the same for
$\Vtgi$ and $\Vrgi$. For $\Om^2 > k^2$ the K-- and I--modes can have
an exponentially growing behaviour whereas for $\Om^2 < k^2$ we use
$\sqrt{\Om^2-k^2} = i\sqrt{|\Om^2-k^2|}$ such that they become
oscillatory. The J-- and Y--modes are always oscillating.

For a given perturbation mode to be physically acceptable one has to
require that, at some initial time $t_\ui$, the perturbations are
small for all values $0<r<\rbr(t_\ui)$, compared to the background. To
check that, we use the limiting forms of the Bessel
functions~\cite{Abramovitz:1970aa}. For large argument, the ordinary
Bessel functions behave as
\begin{equation}
\label{eq:bessellarge}
\begin{aligned}
\besselJ{p}(x) & \underset{x \rightarrow \infty}{\sim}
\sqrt{\dfrac{2}{\pi x}} \cos\left(x - \dfrac{\pi}{2} p -
\dfrac{\pi}{4} \right), \\
\besselY{p}(x) & \underset{x \rightarrow \infty}{\sim}
\sqrt{\dfrac{2}{\pi x}} \sin\left(x - \dfrac{\pi}{2} p -
\dfrac{\pi}{4} \right),
\end{aligned}
\end{equation}
while the modified Bessel functions grow or decrease exponentially
\begin{equation}
\label{eq:Kbessellarge}
\begin{aligned}
\besselK{p}(x) & \underset{x \rightarrow \infty}{\sim}
\sqrt{\dfrac{\pi}{2x}}\ue^{-x},\\
\besselI{p}(x) & \underset{x \rightarrow \infty}{\sim}
\dfrac{1}{\sqrt{2\pi x}}\ue^{x}.
\end{aligned}
\end{equation}
Therefore, in Eqs.~(\ref{eq:bulksolVt}) and (\ref{eq:bulksolVr}) all
modes, except for the K--mode, diverge as $r \ra 0$. Hence the only
regular modes are
\begin{align}
\label{eq:Vtgisol}
\Vtgi &= A(\mathbf{k},\Om) \dfrac{\lbulk^2}{r^2} \besselK{2} \negthickspace
\left(\dfrac{\lbulk^2}{r}\Om \right)
\ue^{\displaystyle \pm t \sqrt{\Om^2-k^2}}, \\
\label{eq:Vrgisol}
\Vrgi &= \pm A(\mathbf{k},\Om) \sqrt{1-\dfrac{k^2}{\Om^2}}
\, \dfrac{\lbulk^2}{r^2} 
\besselK{1} \negthickspace \left(\dfrac{\lbulk^2}{r} \Om\right)
\ue^{\displaystyle \pm t\sqrt{\Om^2-k^2}},
\end{align}
where the amplitude $A(\mathbf{k},\Om)$ is determined by the initial conditions
and carries an implicit spatial index. For small wave numbers,
$k^2<\Om^2$, we are of course only interested in the growing mode but
for large wave numbers both modes are comparable and oscillating in
time. It is easy to see that the K--mode is also normalizable in the
sense that
\begin{equation}
\label{eq:integrability}
\begin{aligned}
\int_0^{\rbr} \sqrt{|g|}|\Vtgi |^2 \ud r &\propto \int_0^{\rbr} \dfrac{1}{r}
\left[\besselK{2} \negthickspace \left(\dfrac{\lbulk^2}{r}\Om \right)
\right]^2 \ud r < \infty, \\
\int_0^{\rbr} \sqrt{|g|}|\Vrgi|^2 \ud r &\propto \int_0^{\rbr}  \dfrac{1}{r}
\left[\besselK{1} \negthickspace \left(\dfrac{\lbulk^2}{r}\Om\right)
\right]^2
\ud r < \infty.
\end{aligned}
\end{equation}
It is important to note at this point that, in this sense, also the
J--modes and Y--modes are normalizable. One might view this
integrability condition as a requirement to insure finiteness of the
energy of these modes. This suggest that the J-- and Y--modes
could also be excited by some physical process. Indeed, from
Eq.~(\ref{eq:adsrs}), their divergence for $r \rightarrow 0$ can be
recast in terms of the $\varrho$ coordinate, with $\varrho \rightarrow
\infty$. Expressed in terms of $\varrho$, the integrability condition
(\ref{eq:integrability}) ensures that the J-- and Y--modes are well
defined in the Dirac sense, and thus that only a superposition of them
may represent physical perturbations\footnote{This is of course not
the case for the I--modes.}. However, at small $r$, even if
integrable, the perturbation amplitudes of the J-- and Y--modes become
large and it is not clear to us in which sense linear gravitational
perturbation theory can remain valid for these modes.

 During a bulk inflationary phase which leads to the
production of $4+1$ dimensional gravity waves, one may expect the
K--modes to be generated, and for a given inflationary model, it
should be possible to calculate the spectrum of fluctuations,
$|A(\mathbf{k},\Om)|^2$. Clearly these fluctuations dominate the
others and are extremely dangerous since exponentially growing modes
could rapidly lead to large fluctuations and destroy the homogeneity
and isotropy of the universe.

Nevertheless, in order to conclude about the viability of these modes,
we first have to derive the evolution of their observational
counterparts on the brane, especially in terms of cosmic time. This is
done in the next section through the perturbed junction conditions.


\section{Gauge invariant perturbation equations on the brane}
\label{sec:branepert}
\subsection{Brane perturbation variables}

Since we are interested in vector perturbations on the brane
induced by those in the bulk, we parameterize the induced
perturbed metric as
\begin{equation}
\label{eq:pertinducedmetric}
\begin{aligned}
\ud \sbrpert^2  & = \qpert_{\mu\nu}\ud y^{\mu} \ud y^{\nu} \\
 & = - \ud \tau^2  + 2 a
\vt_i \ud \tau \ud y^i \\
& + a^2 \left( \delta_{ij} + \nabla_i \vs_i +
\nabla_j \vs_i \right) \ud y^i \ud y^j,
\end{aligned}
\end{equation}
where $\vs^i$ and $\vt^i$ are divergenceless vectors. The junction
conditions which relate the bulk perturbation variables to the
perturbations of the brane can be written in terms of gauge invariant
variables. Under an infinitesimal transformation $y^\mu \rightarrow
y^\mu + \xi^\mu$, where $\xi_\mu=(0,a^2 \xi_i)$, we have
\begin{equation}
\begin{aligned}
\vs_i &\rightarrow \vs_i + \xi_i,\\
\vt_i &\rightarrow \vt_i + a \dot{\xi_i}.
\end{aligned}
\end{equation}
Here the dot is the derivative with respect to the brane time $\tau$.
Hence the gauge invariant vector perturbation is~\cite{Durrer:1993db}
\begin{equation}
\label{eq:defvtgi} \vtgi_i = \vt_i - a \dot{\vs_i}.
\end{equation}
It fully describes the vector metric perturbations on the brane.

The brane energy momentum tensor $S_{\mu \nu}$ given in
Eq. (\ref{eq:branetensor}) has also to be perturbed. As we shall see,
the junction conditions (together with $Z_2$-symmetry) do require a
perturbed energy momentum tensor on the brane. Since we only consider
vector perturbations $\delta \rho = \delta P = 0$. However, the
four-velocity of the perfect fluid does contain a vector perturbation
$\upert^\mu=u^\mu + \delta u^\mu$, with
\begin{equation}
\delta u^\mu =
\begin{pmatrix}
 0 \\ \dfrac{\vel^i}{a},
\end{pmatrix}
\end{equation}
and where $\vel^i$ is divergenceless. Under $y^\mu \rightarrow y^\mu
+ \xi^\mu$,
\begin{equation}
\vel_i \rightarrow \vel_i - a \dot{\xi_i},
\end{equation}
where $\vel_i \equiv \delta_{ij}\vel^i$.
A gauge invariant perturbed velocity can therefore be defined as
\footnote{Alternatively, one may use the fact that the
covariant component $\upert_i$ is already gauge invariant to
find the gauge invariant quantity.}
\begin{equation}
\label{eq:defvelgi} \velgi_i=\vel_i + a \dot{\vs_i}.
\end{equation}
In addition, the anisotropic stresses contain a vector component
denoted $\aniso_i$. Since the
corresponding background quantity vanishes, this variable is
gauge invariant according to the Stewart--Walker
lemma~\cite{Stewart:1984}.

In summary, there are three gauge invariant brane perturbation
variables. We shall use the combinations
\begin{equation}\label{eq:branegi}
\begin{aligned}
    &\vtgi_i = \vt_i - a \dot{\vs_i}, \\
    &\velgi_i=\vel_i + a \dot{\vs_i},\\
    &\aniso_i.
\end{aligned}
\end{equation}

To apply the junction conditions we need to determine the perturbations
of the reduced energy momentum tensor defined in 
Eq.~(\ref{eq:jump}). In terms of our gauge invariant quantities they read
\begin{align}
\label{eq:Spert00}
\delta \widehat{S}_{\tau \tau} & = 0, \\
\label{eq:Spert0i} \delta \widehat{S}_{\tau i} & = -a \left( P +
\frac{2}{3} \rho - \frac{1}{3}
\tension \right) \vtgi_i - a \left(P + \rho \right) \velgi_i,\\
\label{eq:Spertij} \delta \widehat{S}_{ij} & = a^2 P
\left(\partial_i \aniso_j + \partial_j \aniso_i \right).
\end{align}


\subsection{Perturbed induced metric and extrinsic curvature}

We now express the perturbed induced metric, and the perturbed
extrinsic curvature in terms of the bulk perturbation
variables~\cite{Deruelle:2000yj}. In principle there are two
contribution to the brane perturbations: perturbations of the bulk
geometry as well as perturbations of the brane position. A bulk
perturbed quantity has then to be evaluated at the perturbed brane
position [see Eq.~(\ref{eq:embedding})]. Using reparametrization
invariance on the brane~\cite{Deruelle:2000yj}, the latter can be
described by a single variable $\locpert$,
\begin{equation}
\label{eq:pertembedding}
\Xpert^{\si{M}} = X^{\si{M}} + \locpert n^{\si{M}},
\end{equation}
where all quantities are functions of the brane coordinates
$y^\mu$. Since $\locpert$ is a scalar perturbation it does not play a
role in our treatment, and we can consider only the perturbations
$\delta g_{\si{A B}}$ due to the perturbed bulk geometry evaluated at
the unperturbed brane position.
The induced metric perturbation is given by
\begin{equation}
\label{eq:pertqmunu} \delta q_{\mu\nu} = \qpert_{\mu \nu} - q_{\mu
\nu} = e^{\si{A}}_\mu e^{\si{B}}_\nu \delta g_{\si{AB}}.
\end{equation}
From Eqs.~(\ref{eq:bulkpertmetric}), (\ref{eq:defVtgi}),
(\ref{eq:defVrgi}) and (\ref{eq:pertqmunu}) one finds in the gauge
$E_i=0$
\begin{equation}\label{eq:pertqmunucompts}
\begin{aligned}
\delta q_{\tau \tau} &= 0,\\
\delta q_{\tau i} &=a\sqrt{1+\lbulk^2 H^2} \Vtgi_i + a\lbulk H \Vrgi_i,\\
\delta q_{i j} &=\delta g\ei\jei =0.
\end{aligned}
\end{equation}
The time component vanishes as it is a pure scalar, and the purely
spatial components can be set to zero without loss of generality by
gauge fixing.

In the same way, perturbing Eq.~(\ref{eq:extusefull}), and making
use of Eqs.~(\ref{eq:tangent}), (\ref{eq:normal})
in order to derive the perturbed normal vector, leads to (again we use
the gauge $E_i=0$)
\begin{align}
\delta K_{\tau \tau} &= 0  \\
\label{eq:Kpert0i}
\delta K_{\tau i} & = \dfrac{1}{2} \partial_t \Vrgi_i - \dfrac{1}{2} a^2
\partial_r \Vtgi_i -a H \sqrt{1 +\llhh} \Vrgi_i \nonumber \\
& - \dfrac{a}{\lbulk} \left(1 + \llhh\right) \Vtgi_i,\\
\label{eq:Kpertij}
\delta K_{ij} &=  \dfrac{1}{2} a \lbulk H
\left(\partial_i \Vtgi_j + \partial_j \Vtgi_i \right) \nonumber \\
&+ \dfrac{1}{2} a \sqrt{1+ \llhh} \left( \partial_i \Vrgi_j +
\partial_j \Vrgi_i \right),
\end{align}
where $\delta K_{\mu \nu} = \Kpert_{\mu \nu} - K_{\mu \nu}$, and all
bulk quantities have to be evaluated at the brane position. In the
derivation we have also used that on the brane $\partial_\mu =
e^{\si{A}}_\mu \partial_{\si{A}}$.


\subsection{Perturbed junction conditions and solutions}
\label{subsec:perturbedjunction}
The first junction condition requires the first fundamental form
$q_{\si{AB}}$ to be continuous across the brane. Therefore, the
components of the induced metric (\ref{eq:pertinducedmetric}) are
given by the explicit expressions (\ref{eq:pertqmunucompts}). This
leads to the following relations
\begin{equation}
\begin{aligned}
\vs_i & = \Vs_i ,\\
\vt_i & = \sqrt{1+ \llhh} \Vt_i + \lbulk H \Vr_i,
\end{aligned}
\end{equation}
where the bulk quantities have to be taken at the brane position
$(\tbr,\rbr)$. For $\sig_i=\vt_i -a\dot\vs_i$ we use 
\begin{equation}
\begin{aligned}
a\dot\vs_i &= a \left(\dot\tbr \dd_t \Vs_i  + \dot\rbr \dd_r \Vs_i \right) \\
 & = \sqrt{1+ \llhh}\dd_t\Vs_i +a^2 \lbulk H\dd_r \Vs_i.
\end{aligned}
\end{equation}
Together with Eqs.~(\ref{eq:defVtgi}) and (\ref{eq:defVrgi}) this
gives
\begin{equation}
\label{eq:link} \vtgi_i = \sqrt{1+ \llhh} \Vtgi_i + \lbulk H
\Vrgi_i.
\end{equation}
The equations corresponding to the second junction condition are
obtained by perturbing Eq.~(\ref{eq:jump}) (using $K^{>}_{\mu\nu} =
-K^{<}_{\mu\nu} = -K_{\mu\nu}$) and inserting the expressions
(\ref{eq:Kpert0i}), (\ref{eq:Kpertij}) for the perturbed extrinsic
curvature tensor, with Eqs.~(\ref{eq:Spert0i}), (\ref{eq:Spertij}) for
the perturbed energy-momentum tensor on the brane. After some algebra
one obtains, respectively, for the $(0i)$ components, and
the  $(ij)$ components respectively
\begin{align}
    \label{eq:vortlink}
    \frac{2 \lbulk \dot{H}}{\sqrt{1 + \llhh}} a
    \left(\vtgi_i + \velgi_i \right) &= a^2 \partial_r \Vtgi_i -
    \partial_t \Vrgi_i, \\
    \label{eq:anisolink}
    \kappareduct^2 a P \aniso_i = -\lbulk H \Vtgi_i &- \sqrt{1+\lbulk^2
    H^2}\Vrgi_i,
\end{align}
where we have used the unperturbed junction conditions,
Eqs.~(\ref{eq:friedmanT}) and (\ref{eq:friedmanS}), and the fact
that on the brane $\partial_\mu = e^{\si{A}}_\mu
\partial_{\si{A}}$. One can use Eq.~(\ref{eq:link}) to eliminate
the variable $\Vrgi_i$ in Eq.~(\ref{eq:anisolink}) in favor of
$\vtgi_i$. This leads to
\begin{equation}\label{eq:anisovectlink}
    \kappareduct^2 a \lbulk H P \aniso_i + \sqrt{1 + \llhh} \vtgi_i  =
    \Vtgi_i.
\end{equation}

Hence, if by some mechanism, like \EG bulk inflation, gravity waves
are produced in the bulk, their vector parts $\Vtgi_i$ and $\Vrgi_i$
will induce vector perturbations $\vtgi_i$, vorticity $\vtgi_i +
\velgi_i$, and anisotropic stresses $\aniso_i$ on the brane according
to Eqs.~(\ref{eq:link}), (\ref{eq:vortlink}) and
(\ref{eq:anisolink}). These are necessary to satisfy the boundary
conditions implied by $Z_2$ symmetry. In general, $\vtgi_i$, $\vtgi_i
+ \velgi_i$, and $\aniso_i$ come from both, bulk perturbations and
intrinsic perturbations on the brane, but here, we want to study only
the perturbations induced from bulk gravity waves and do not consider
additional sources on the brane. Then the brane variables are fully
determined by $\Vtgi_i$ and $\Vrgi_i$.

In the following we do not want to specify a particular mechanism
which generates $\Vtgi_i$ and $\Vrgi_i$, and just assume they have
been produced with some spectrum given by $A({\bf k},\Om)$. In
4-dimensional standard cosmology it is well known that vector
perturbations decay, and therefore they are not considered. Note that
in simple inflationary models they are not even generated. Only if
they are continuously re-generated like, \EG in models with
topological defects (see \EG Ref.~\cite{Durrer:2001cg}), they affect
CMB anisotropies. Here the situation is different since the modes
considered are either exponentially growing or oscillating at least
with respect to the bulk time. To calculate the CMB anisotropies from
the vector perturbations induced by bulk gravity waves, the relevant
quantities are $\vtgi$ and $\velgi +\vtgi$ given in terms of the bulk
variables by Eqs.~(\ref{eq:link}) and (\ref{eq:vortlink}). Inserting
the solutions (\ref{eq:Vtgisol}) and (\ref{eq:Vrgisol}) for the
K--mode into (\ref{eq:link}) and (\ref{eq:vortlink}) yields
\begin{widetext}
\begin{align}
\label{eq:vtgisol}
\vtgi(\tbr,\mathbf{k}) & = A(\mathbf{k},\Om)\left[\sqrt{1+ \llhh}
 \dfrac{1}{a^2} \besselK{2} \negthickspace \left(\dfrac{\lbulk \Om}{a}
 \right) \mp \lbulk H\sqrt{1-\dfrac{k^2}{\Om^2}} \, \dfrac{1}{a^2}
 \besselK{1}
 \negthickspace \left(\dfrac{\lbulk \Om}{a}\right) \right]
\ue^{\displaystyle \pm t_b\sqrt{\Om^2-k^2}}, \\
\label{eq:vortsol}
\left(\vtgi + \velgi \right)(\tbr,\mathbf{k}) & = - A(\mathbf{k},\Om)
 \dfrac{k^2}{\Om^2} \dfrac{\sqrt{1 + \llhh}}{2 \lbulk^2
 \dot{H}}\dfrac{\lbulk \Om}{a} \dfrac{1}{a^2} \besselK{1}
 \negthickspace \left(\dfrac{\lbulk \Om}{a}\right) \ue^{\displaystyle \pm
 t_b\sqrt{\Om^2-k^2}},
\end{align}
\end{widetext}
where again we have omitted the spatial index $i$ on $\vtgi$, $\velgi$
and $A$.
Similar equations can be obtained for the J-- and Y--modes by
replacing, in Eqs.~(\ref{eq:vtgisol}) and (\ref{eq:vortsol}), the
modified Bessel function by the ordinary ones, plus the
transformations: $-k^2 \rightarrow k^2$, $H \rightarrow -H$ and $t
\rightarrow it$.

These equations are still written in bulk time $\tbr$ which is related
to the conformal time $\eta$ on the brane by
\begin{equation}\label{eq:conformaltime}
\ud \tbr = \sqrt{1+ \llhh }\, \ud \eta.
\end{equation}
Therefore, at sufficiently late time $\llhh \ll 1$ such that $\ud \tbr
\simeq \ud\eta$. Note that $\lbulk$ is the size of the extra-dimension
which must be smaller than micrometers while $H^{-1}$ is the Hubble
scale which is larger than $10^5$ light years at times later than
recombination which are of interest for CMB anisotropies.

As a result, the growing or oscillating behaviour in bulk time remains
so in conformal time. Moreover there are additional time dependent
terms in Eqs.~(\ref{eq:vtgisol}) and (\ref{eq:vortsol}) with respect
to Eqs.~(\ref{eq:Vtgisol}) and (\ref{eq:Vrgisol}) due to the motion of
the brane. As can be seen from Eqs.~(\ref{eq:vtgisol}) and
(\ref{eq:vortsol}), the modes evolve quite differently for different
values of their physical bulk wave number $\Om/a$. In the limit
$\Om/a \ll 1/\lbulk$ and for $\Om^2>k^2$, the growing K-modes behave like
\begin{equation}
\label{eq:Ksolsmall}
\begin{aligned}
\vtgi & \sim \dfrac{2 A}{(\Om \lbulk)^2}
\,\ue^{ \eta \sqrt{\Om^2-k^2}}, \\ \vtgi + \velgi & \sim
-\dfrac{A}{(\Om \lbulk)^2} \dfrac{k^2}{2 a^2 \dot{H}}
\,\ue^{ \eta \sqrt{\Om^2-k^2}},
\end{aligned}
\end{equation}
where use has been made of $\llhh \ll 1$, and of the limiting forms of
Bessel function for small arguments~\cite{Abramovitz:1970aa}
\begin{equation}
\besselK{p}(x) \underset{x \rightarrow 0}{\sim} \dfrac{1}{2} \Gamma(p)
\left(\dfrac{2}{x}\right)^p.
\end{equation}
In the same way, from Eq.~(\ref{eq:Kbessellarge}), the K--modes
verifying $\Om/a \gg 1/\lbulk$ reduce to
\begin{equation}
\label{eq:Ksollarge}
\begin{aligned}
\vtgi & \sim \dfrac{A}{(\Om \lbulk)^2}\ue^{ \eta
\sqrt{\Om^2-k^2}} \sqrt{\dfrac{\pi}{2}} \left(\dfrac{\Om
\lbulk}{a}\right)^{3/2} \ue^{ - \Om\lbulk/a}, \\ \vtgi +
\velgi & \sim -\dfrac{A}{(\Om \lbulk)^2}\dfrac{k^2}{2 a^2
\dot{H}}\,\ue^{ \eta \sqrt{\Om^2-k^2}}
\sqrt{\dfrac{\pi}{2}} \left(\dfrac{\Om \lbulk}{a}\right)^{1/2}
\ue^{ - \Om\lbulk/a}.
\end{aligned}
\end{equation}
They are exponentially damped compared to the former [see
Eq.~(\ref{eq:Ksolsmall})]. As a result, the main contribution of the
K-mode vector perturbations comes from the modes with a physical wave
number $\Om/a$ smaller than the energy scale $1/\lbulk$ associated
with the extra-dimension. As the universe expands, a mode with fixed
value $\Om$ remains relatively small as long as the exponents in
Eq.~(\ref{eq:Ksollarge}) satisfy
\begin{equation}
\dfrac{\Om}{a} L - \eta\sqrt{\Om^2-k^2} \simeq \dfrac{1}{a} \left(\Om
\lbulk -\tau\sqrt{\Om^2-k^2}\right) > 0.
\end{equation}
When this inequality is violated, for $k\ll\Om$ this is soon
after $\tau\sim \lbulk$, the mode starts growing exponentially.
The time $\tau\sim\lbulk$ also corresponds to the initial time at
which the evolution of the universe starts to become Friedmannian.

In the same way, one can derive the behaviors of the J-- and Y--modes
on the brane for physical bulk wave numbers greater or smaller than
the size of the extra-dimension. This time, the exponentially growing
terms are replaced by oscillatory ones, and the ordinary Bessel
functions are approximated by \{see Eq.~(\ref{eq:bessellarge})
and~\cite{Abramovitz:1970aa}\}
\begin{equation}
\label{eq:besselsmall}
\begin{aligned}
\besselJ{p}(x) &\underset{x \rightarrow 0}{\sim}
\dfrac{1}{\Gamma(p+1)} \left(\dfrac{x}{2} \right)^p,\\
\besselY{p}(x) &\underset{x \rightarrow 0}{\sim}
- \dfrac{1}{\pi} \Gamma(p) \left(\dfrac{2}{x} \right)^p.
\end{aligned}
\end{equation}

From Eqs.~(\ref{eq:bessellarge}) and (\ref{eq:besselsmall}), the
equivalents of Eqs.~(\ref{eq:vtgisol}) and (\ref{eq:vortsol}) for J--
and Y--modes can be shown to oscillate always. From
Eq.~(\ref{eq:nowhubbleevol}), their amplitude is found to decay like
$a^{-3/2}$ in the short wavelength limit $\Om/a \gg 1/\lbulk$. In
the long wavelength limit $\Om/a \ll 1/\lbulk$, the amplitude of
the Y-mode stays constant whereas the J-mode decreases as $a^{-4}$.

The vorticity is also found to oscillate in conformal time. This time,
the amplitude of the long wavelength Y--modes always grows as
$a^{3w+1}$ while for the J--modes it behaves like $a^{3w-1}$. Finally,
in the short wavelength limit, both Y and J vorticity modes grow
like $a^{3w+1/2}$.

Whatever the kind of physical vector perturbation modes excited in the
bulk, we have shown that there always exist bulk wave numbers $\Om$
that give rise to growing vector perturbations on the brane. Although
the J-- and Y--modes generate vorticity growing like a power law of
the scale factor, they can be, in a first approximation, neglected
compared to the K--modes which grows like an exponential of the
conformal time. In the next section, we will derive constraints
on the amplitude $A(\mathbf{k},\Om)$ of the K--modes by 
estimating the CMB anisotropies they induce.


\section{CMB anisotropies}
\label{sec:cmb}

In order to determine the temperature fluctuations in the CMB due to
vector perturbations on the brane, we have to calculate how a photon
emitted on the last scattering surface travels through the perturbed
geometry (\ref{eq:pertinducedmetric}). A receiver today therefore
measures different microwave background temperatures $T_{\si\uR}(n^i)$
for incident photons coming from different directions $n^i$. In terms
of conformal time the vector-type temperature fluctuations are given
by~\cite{Durrer:1993db}
\begin{equation}\label{eq:tempfluctuations}
\begin{aligned} 
\dfrac{\delta T_{\si \uR}(n^i)}{T_{\si \uR}} &= n ^i \left.(\vtgi_i +
  \velgi_i)\right|^{\si\uR}_{\si\uE} +
  \int^{\si\uR}_{\si\uE}\dfrac{\partial \vtgi_i}{\partial x^j} n^i n^j
  \ud\lambda,\\
 &= -n^i\velgi_i(\etae) + \int^{\si\uR}_{\si\uE}\vtgi_i' n^i 
  \ud \eta
\end{aligned}
\end{equation}
where $\lambda$ denotes the affine parameter along the photon
trajectory and the prime is a derivative with respect to conformal
time $\eta$. The ``R'' and ``E'' index refer to the time of photon
reception and emission, which in the following correspond to the
conformal times today $\etao$ and at recombination $\etae$. For the
second equality we have used
\begin{equation}
\dfrac{\ud \vtgi_i}{\ud \lambda} =\vtgi_i' - 
      n^j\dfrac{\partial \vtgi_i}{\partial x^j},
\end{equation}
where $-n^i$ is the direction of the photon momentum. We have also
neglected the contribution from the upper boundary, ``R'', in the
first term since it simply gives rise to a dipole term. The first
term in Eq.~(\ref{eq:tempfluctuations}) is a Doppler shift, and the
second is known as integrated Sachs-Wolfe effect. To determine the
angular CMB perturbation spectrum $\Cl$, we apply the total angular
momentum formalism developed by Hu and
White~\cite{Hu:1997hp}. According to this, a vector perturbation
$\mathbf{v}$ is decomposed as
\begin{equation}
\mathbf{v} = \mathbf{e}^+ v^+ + \mathbf{e}^-v^-,
\end{equation}
where
\begin{equation}
\mathbf{e}^\pm = \dfrac{-i}{\sqrt{2}}\left({\bf e}^{(1)} \pm i{\bf
 e}^{(2)}\right),
\end{equation}
and $\mathbf{e}^{(1,2)}$ are defined so that $(\mathbf{e}^{(1)}$,
$\mathbf{e}^{(2)}$, $\hat{\mathbf{k}}=\mathbf{k}/k)$ form a
righthanded orthonormal system. Using this decomposition for 
$\velgi_i$ and $\vtgi_i$, one obtains the angular CMB perturbation
spectrum $\Cl$ via
\begin{equation}
\label{eq:Cl}
\Cl = \dfrac{2}{\pi}\ell(\ell+1) \int_0^\infty
 k^2 \langle|\De_\ell(k)|^2\rangle \ud k
\end{equation}
where
\begin{equation}
\label{eq:Del}
\begin{aligned}
\De_\ell(k) &= - \velgi^+(\etae,k) \dfrac{j_\ell(k \etao-
k\etae)}{k \etao - k\etae} \\ & + \int_{\etae}^{\etao} \vtgi^{+\prime}
(\eta,k) \dfrac{j_\ell(k\etao-k\eta)}{k\etao-k\eta}\ud\eta.
\end{aligned}
\end{equation}
In Eq.~(\ref{eq:Del}) we have assumed that the process which generates
the fluctuations has no preferred handedness so that
$\langle|\vtgi^+|^2\rangle =\langle|\vtgi^-|^2\rangle$ as well as
$\langle|\velgi^+|^2\rangle =\langle|\velgi^-|^2\rangle$. Omitting the
``$\pm$'' superscripts, we can take into account the negative helicity
mode simply by a factor $2$.

As shown in the previous section, the main contribution of the
K--modes comes from those having long wavelengths $a/\Om \gg \lbulk$,
and $k < \Om$. In the following, only these modes will be
considered. Since they are growing exponentially in $\eta$, the
integrated Sachs-Wolfe contribution will dominate and we concentrate
on it in what follows. A more rigorous justification is given in the
appendix. Inserting the limiting form (\ref{eq:Ksolsmall}) for $\vtgi$
in Eq.~(\ref{eq:Del}) gives
\begin{equation}
\label{eq:exp}
\begin{aligned}
\De_\ell(\kb) & \simeq   2 \Ao \Om^{n} \kb^n
\ue^{ \vo \sqrt{1-\kb^2}} \sqrt{\dfrac{1}{\kb^2}-1~} \\
& \times \int_{0}^{\xe}
\dfrac{j_\ell(\xx)}{\xx} \ue^{ - \xx \sqrt{1/\kb^{2}-1}}
\ud \xx, 
\end{aligned}
\end{equation}
where a simple power law ansatz has been chosen for the primordial
amplitude
\begin{equation}
\sqrt{\langle \left|A(\mathbf{k},\Om)\right|^2 \rangle} = \Ao(\Om)
\Om^2 \lbulk^2 k^n.
\end{equation}
A dimensionless wave number $\kb$, and conformal time $\vv$, have also
been introduced as
\begin{equation}
\kb = \dfrac{k}{\Om}, \quad \vv = \eta \Om,
\end{equation}
in order to measure their physical counterparts in unit of the bulk
wavelengths. The condition $k < \Om$ becomes now $\kb <
1$. The integration over $\eta$ in the integrated Sachs-Wolfe term is
now transformed into an integration over the dimensionless variable
$x$ defined by 
\begin{equation}
x = k \left(\etao - \eta \right) = \kb \left(\vo - \vv \right).
\end{equation}
In the following, we will use the approximation $\xe = k \left(\etao -
\etae \right) \simeq k\etao$.

By observing the CMB, one may naturally expect that the
perturbations with physical wavelength greater than the horizon size
today have almost no effect. In terms of our parameters, this means that
the main contribution in the $\Cl$ comes from the modes
verifying $\Omega/\az > \Ho$, hence $\vo >1$.

In the appendix, we derive a crude approximation for the angular power
spectrum induced by the exponentially growing K--modes, in a range a
little more constrained than the one previously motivated, namely
\begin{equation}
\label{eq:waverange}
\ell_{\max} \Ho<\dfrac{\Om}{\az} < \dfrac{\lbulk^{-1}}{1+\ze},
\end{equation}
where $\ze$ is the redshift at photon emission which is taken to
coincide with recombination, $\ze\simeq 10^3$. In order to simplify
the calculation, we do not want the transition between the damped
K--modes ($\Om/a >\lbulk^{-1}$) and the exponentially growing ones
($\Om/a <\lbulk^{-1}$) to occur between the last scattering surface
and today.  This requirement leads to the upper limit of
Eq.~(\ref{eq:waverange}). Moreover, in order to derive the $\Cl$, we
perform an expansion with respect to a parameter $\ell/\vo$ assumed
small, and $\ell_{\max}$ refers to the multipole at which this
approximation breaks down. The lower limit in Eq.~(\ref{eq:waverange})
comes from this approximation. Using the values $\lbulk \simeq 10^{-3}
\mm$, $\Ho^{-1} \simeq 10^{29} \mm$, $\ell_{\max} \simeq 10^{3}$, and
$\ze \simeq 10^{3}$, one finds
\begin{equation}
\label{eq:numwaverange}
10^{-26} \mm^{-1}<\dfrac{\Om}{\az} < 1 \mm^{-1}.
\end{equation}
The corresponding allowed range for the parameter $\vo$ becomes [see
Eq.~(\ref{eq:defvove})]
\begin{equation}
\label{eq:vorange}
10^{3} < \vo < 10^{29}.
\end{equation}

Clearly the detailed peak structure on the CMB anisotropy spectrum
would have been different if we had taken into account the oscillatory
parts ($k>\Om$) of the K--modes, as well as the Y-- and J--modes, but
here we are only interested in estimating an order of magnitude
bound. As detailed in the appendix, for a scale invariant initial
spectrum, \IE $n=-3/2$, we obtain
\begin{equation}
\dfrac{\ell(\ell+1)}{2\pi}\Cl \gtrsim \left(\Ao \ue^{\vo}\right)^2
\dfrac{\ue^{-\ell}}{\ell^{7/2}} \left(\dfrac{\ell}{\vo}
\right)^{\ell -1}.
\end{equation}
From current observations of the CMB
anisotropies, the left hand side of this expression is about
$10^{-10}$, and for $\ell\simeq 10$, one gets
\begin{equation}
\Ao(\Om) \lesssim \dfrac{\ue^{-\left[\vo - 5 \ln(\vo)\right]}}{10^5}.
\end{equation}
From Eq.~(\ref{eq:defvove}) and (\ref{eq:vorange}), one find
that the primordial amplitude of these modes must satisfy
\begin{equation}
\label{eq:weakconstraint}
\Ao(\Om)  < \ue^{-10^{3}}, \textrm{ for } \Om/\az \simeq 10^{-26}\mm^{-1}
\end{equation}
 and, more dramatically,
\begin{equation}
\label{eq:strongconstraint}
\Ao(\Om)  < \ue^{-10^{29}},   \textrm{ for } \Om/\az \simeq 1 \mm^{-1}.
\end{equation}
As expected, the perturbations with wavelength closer to the horizon
today (smaller values of $\Om$) are less constrained than smaller
wavelengths [see Eq.~(\ref{eq:weakconstraint})]. Moreover, one may
expect that the bound (\ref{eq:strongconstraint}) is no longer valid
for $\Om/\az > \lbulk^{-1}/(1+\ze)$ since the modes in
Eq.~(\ref{eq:Ksollarge}) start to contribute. However, the present
results concern more than $20$ orders of magnitude for the physical
bulk wave numbers $\Om/\az$, and show that the exhibited modes are
very dangerous for braneworlds.

It seems that the only way to avoid these constraints is to find a
physical mechanism forbidding any excitation of these modes.


\section{Conclusion}

In this paper we have shown that vector perturbations in the bulk
generically lead to growing vector perturbations on the brane in the
Friedmann-Lema\^{\i}tre era. This behaviour radically differs from the
usual one in four-dimensional cosmology, where vector modes decay like
$a^{-2}$ whatever the initial conditions.

Among the growing modes, we have identified modes which are
perfectly normalizable and lead to exponentially growing vector
perturbations on the brane with respect to conformal time. By
means of a rough estimate of the CMB anisotropies induced by these
perturbations, we have found that they are severely
incompatible with a homogeneous and isotropic universe by lighting up
a fire in the microwave sky, unless their primordial amplitude is
extremely small.
 
No particular mechanism for the generation of these modes has been
specified. However, we may expect that bulk inflation leads to
gravitational waves in the bulk which do generically contain
them. Even if they are not generated directly, they should be induced
in the bulk by second order effects. Usually, these effects are too
small to have any physical consequences, but here they would largely
suffice due to their exponential growth [see
Eqs.~(\ref{eq:weakconstraint}) and (\ref{eq:strongconstraint})]. This
second order induction seems very difficult to prevent in the models
discussed here.

It is interesting to note that this result is also linked to the
presence of a non-compact extra-dimension which allows a continuum in
the bulk frequency modes $\Om$. A closer examination of
Eq.~(\ref{eq:VtgiT}) shows that the mode $\Om=0$, admits only J-- and
Y--mode behaviours. In a compact space, provided the first quantized
value of $\Om$ is sufficiently large, one could expect the
exponentially growing K--modes to
be never excited by physical processes. Another more speculative way
to dispose of them could be to study their causal structure: By
setting $k=0$ in Eq.~(\ref{eq:VtgiT}), one might be tempted to view
$-\Om^2$ as a negative mass squared, suggesting that these K--modes
could physically represent bulk tachyons. This would motivate their
causal decoupling from the subsequent evolution of the brane. In a
more basic theory, which goes beyond our classical relativistic
approach, these modes may thus not be present at all. However, this
interpretation breaks down for modes with $k > \Om$ which can
again influence the brane light cone.

It is important to retain that even if the K--modes can be eliminated
in some way, the growing behavior of the Y-- and J--modes
remains. Although their power law growth is not as critical as the
exponential growth of the K--modes, they should have significant
effects on the CMB anisotropies. Indeed, they lead to amplified
oscillating vector perturbations which are entirely absent in
four-dimensional cosmology. This will be the object of a future
study~\cite{timon:2003}.

We therefore conclude that, if no physical mechanism forbids the
generation of the discussed vector modes with time dependence $\propto
\exp(\eta\sqrt{\Om^2-k^2})$, anti-de Sitter braneworlds, with
non-compact extra-dimension cannot reasonably lead to a
homogeneous and isotropic expanding universe.

\acknowledgments It is a pleasure to thank C.~Deffayet and F.~Vernizzi
for enlightening discussions. This work is supported by the Swiss
National Science Foundation. We also acknowledge technical and moral
support by Martin Zimmermann in the last phase of the project.

\appendix*
\begin{widetext}
\section{CMB angular power spectrum}
In this appendix we first present a crude and then a more
sophisticated approximation for the $C_\ell$--power spectrum from the
exponentially growing K--modes. As we shall see, at moderate values of
$\ell\sim$ $10$--$50$, both lead to roughly the same bounds for the amplitudes
which are also presented in the text.

\subsection{Crude approximation}
Here we start from Eq.~(\ref{eq:exp}). In the integral
\begin{equation}
\int_{0}^{\xe} \dfrac{j_\ell(\xx)}{\xx} \ue^{ - \xx \sqrt{1/\kb^{2}-1}}
\ud \xx,
\end{equation}
we replace $j_\ell$ by its asymptotic expansion for small $\ell$,
\begin{equation}
j_\ell(x)\simeq
\left(\dfrac{x}{2}\right)^\ell\dfrac{\sqrt{\pi}}{2\Gamma(\ell+3/2)}.
\end{equation}
This is a good approximation if either $\xe \simeq k\eta_0<\ell/2$ or
$\xe(1/\kb^2-1)^{1/2}>1$. The integral of $x$ then gives
\begin{equation}
\label{eq:Delcrude}
\langle|\De_\ell(\kb)|^2\rangle \simeq \dfrac{\pi
\Ao^2 \Om^{2n} \kb^{2n} }{2^{2\ell} \ell^{3}}
\left(\dfrac{1}{\kb^2}-1\right)^{1-\ell} \ue^{ 2\vo \sqrt{1-\kb^2}}.
\end{equation}
Integrating over $k$, we must take into account that our approximation
is only valid for $k<k_{\max}= (\Om^2-\eta_0^{-2})^{1/2}$. Since we
integrate a positive quantity we certainly obtain a lower bound by
integrating it only until $k_{\max}$. To simplify the integral we also
make the substitution $y=\sqrt{1-\kb^2}$. With this, and inserting our
result (\ref{eq:Delcrude}) in Eq.~(\ref{eq:Cl}), we obtain
\begin{equation}
\label{eq:Clcrude} \ell^2C_\ell \gtrsim \dfrac{2\ell}{2^{2\ell}}
\Ao^2\Om^{2n+3}\int_{1/\vo}^1(1-y^2)^{n+\ell-1/2} y^{3-2\ell}e^{2\vo
y}\ud y . 
\end{equation}
For $\ell\ge 2$, $ y^{3-2\ell}\ge 1$ on the entire range of
integration. Hence we have
\begin{equation} \label{eq:Clcrude1}
\ell^2C_\ell \gtrsim
\dfrac{2\ell}{2^{2\ell}}\Ao^2\Om^{2n+3}\int_{0}^1(1-y^2)^{n+\ell-1/2}
\ue^{2\vo y} \ud y.
\end{equation}
This integral can be expressed in terms of modified Struve
functions~\cite{Abramovitz:1970aa}. In the interesting range, $\vo\gg
1$ we have
\begin{equation}
\int_{0}^1(1-y^2)^{n+\ell-1/2} \ue^{2\vo y}dy \simeq \ue^{2\vo}
    \dfrac{\Gamma(n+\ell+1/2)}{2\vo^{n+\ell+1/2}}.
\end{equation}
Inserting this result in Eq~(\ref{eq:Clcrude1}) we then finally obtain
\begin{equation} 
\label{eq:Clcruden}
\begin{aligned}   
\ell^2C_\ell \gtrsim \sqrt{2\pi} \Ao^2\Om^{2n+3} \ue^{2\vo} \dfrac{
  \sqrt{\ell}\, \ue^{-\ell} } {2^{2\ell}}
  \left(\dfrac{\ell}{\vo}\right)^{n+\ell+1/2} = \sqrt{2\pi} \Ao^2 \ue^{2\vo}
  \dfrac{ \sqrt{\ell}\, \ue^{-\ell}}{2^{2\ell}}
  \left(\dfrac{\ell}{\vo}\right)^{\ell-1},
\end{aligned}
\end{equation}
where we have used Stirling's formula for $\Gamma(\ell+n+1/2)$ and set
$n=-3/2$ after the equality sign.

In the next section we use a somewhat more sophisticated method which
allows us to calculate also the Doppler contribution to the
$C_\ell$'s. For the ISW effect this method gives
\be \label{eq:ISWsoph}
\ell^2C_\ell \simeq \sqrt{\dfrac{2}{\pi}}\dfrac{e^{-\ell}}{36\ell^{7/2}}
  \left(\dfrac{\ell}{\vo}\right)^{\ell-1} 
   \Ao^2 \ue^{2\vo}
\ee
for $n=-3/2$. Until $\ell \sim 15$ the two approximations are in
reasonable agreement and lead to the same prohibitive bounds for
$\Ao(\Om)$. For $\ell>15$, Eq.~(\ref{eq:ISWsoph}) becomes more stringent.

\subsection{Sophisticated approximation}
In Eq.~(\ref{eq:exp}) we have only considered the dominant
contribution coming from the integrated Sachs-Wolfe effect. The
general expression is obtained by inserting the solutions
(\ref{eq:Ksolsmall}) for $\vtgi$ and $ \velgi$ in Eq.~(\ref{eq:Del}),
\begin{equation}
\label{eq:expAp}
\begin{aligned}
\De_\ell(\kb) &= 2\Ao \Om^n \kb^n \left(1 -
\dfrac{\kb^2}{\kbe^2}\right) \dfrac{j_\ell(\kb \vo- \kb\ve)}{\kb \vo -
\kb\ve} \ue^{ \ve \sqrt{1-\kb^2}} + 2 \Ao \Om^{n} \kb^n
\ue^{ \vo \sqrt{1-\kb^2}} \sqrt{\dfrac{1}{\kb^2}-1~}  \int_{0}^{\xe}
\dfrac{j_\ell(\xx)}{\xx} \ue^{ - \xx \sqrt{1/\kb^{2}-1}}
\ud \xx. \\
\end{aligned}
\end{equation}
To derive the first term we have used Eq.~(\ref{eq:nowhubbleevol}) in
the matter era. The parameter
\begin{equation}
\kbe^2 = 6 (1 + \ze)  \left(\dfrac{\Ho\az}{\Om}\right)^2,
\end{equation}
reflects the change in behavior of the modes, redshifted by $\ze$ to
the emission time, which are either outside or inside the horizon
today. It is important to note that the parameter $\Ho \az/\Om$
completely determines the effect of the bulk vector perturbations on
the CMB, together with the primordial amplitude $\Ao$. Indeed, solving
Eq.~(\ref{eq:hubbleevol}) in terms of conformal time, and using
Eqs.~(\ref{eq:enercons}) and (\ref{eq:nowhubbleevol}), yields $\etao
\simeq 2/(\az \Ho)$ in the Friedmann-Lema\^{\i}tre era. Thus
\begin{equation}
\label{eq:defvove}
\vo \simeq 2 \dfrac{\Om/\az}{\Ho}, \quad \ve \simeq \dfrac{1}{1 + \ze}
\dfrac{\Om/\az}{\Ho}.
\end{equation}
We now replace the spherical Bessel functions $\besselj{\ell}$ in the
integrated Sachs-Wolfe term (ISW) using the
relation~\cite{Abramovitz:1970aa}
\begin{equation}\label{}
    \besselj{\ell}(x)=
    \sqrt{\frac{\pi}{2x}}\besselJ{\ell+1/2}(x).
\end{equation}
In the ISW term the upper integration limit can be taken to be
infinity as the contribution from $x_{\Esi}$ to infinity can be
neglected provided $x \sqrt{1/\kb^2-1}>1$. This restriction is
equivalent to $\kb^2 < 1-1/\vo$ which is verified for almost all
values of $\kb$ up to one, given that $\vo$ varies in the assumed
range (\ref{eq:vorange}). We remind that for the exponentially growing
K-mode $k \leq \Omega$, and hence $0 \leq \bar{k} \leq 1$. This allows
for the exact solution~\cite{Gradshteyn:1965aa}
\begin{equation}\label{eq:DeltaISWF}
    \int_0^\infty x^{-3/2}\besselJ{\ell+1/2}(x)
    e^{-x\sqrt{1/\kb^2-1}}\ud x
    = \frac{\kb^\ell}{2^{\ell+1/2}}\frac{\Gamma(\ell)}{\Gamma(\ell+3/2)}
    F\left(\frac{\ell}{2},\frac{\ell}{2}+1;\ell+\frac{3}{2};\bar{k}^2\right),
\end{equation}
where $F$ is the Gauss hypergeometric function. In regard to the
subsequent integration over $k$ we approximate $F$ as follows. For
small values of $\kb$, $F$ is nearly constant with value $1$, at
$\kb=0$. As $\kb \rightarrow 1$ the slope of $F$ diverges and it
cannot be Taylor expanded anymore. However, by means of the linear
transformation formulas~\cite{Abramovitz:1970aa}, $F$ can be written
as a combination of hypergeometric functions depending on
$1-\bar{k}^2$
\begin{equation}
\begin{aligned}
F\left(\dfrac{\ell}{2},\dfrac{\ell}{2}+1;\ell +
\dfrac{3}{2};\kb^2\right) & = \dfrac{\Gamma\left(\ell +
\dfrac{3}{2}\right)
\Gamma\left(\dfrac{1}{2}\right)}{\Gamma\left(\dfrac{\ell}{2} +
\dfrac{3}{2}\right) \Gamma\left(\dfrac{\ell}{2} + \dfrac{1}{2}\right)}
F\left(\dfrac{\ell}{2},\dfrac{\ell}{2}+1; \dfrac{1}{2};1-\kb^2\right)
\\ & + \sqrt{1-\kb^2} \dfrac{\Gamma\left(\ell + \dfrac{3}{2}\right)
\Gamma\left(-\dfrac{1}{2}\right)}{\Gamma\left(\dfrac{\ell}{2}\right)
\Gamma\left(\dfrac{\ell}{2} + 1\right)} F\left(\dfrac{\ell}{2} +
\dfrac{3}{2},\dfrac{\ell}{2}+ \dfrac{1}{2}; \dfrac{3}{2};1-\kb^2\right).
\end{aligned}
\end{equation}
These in turn can be expanded around $1-\kb^2=0$, and gives
\begin{equation}
\label{eq:fone}
F \underset{\kb \rightarrow 1}{\sim} 2^{\ell+1/2} \left(1 - \ell
\sqrt{1-\kb^2} \right).
\end{equation}
These two approximations intersect at $\kb=\sqrt{1-1/\ell^2}$. In this
way, we can evaluate the mean value of $F$ by integrating the two
parts over the interval $[0,1]$. Thus, the hypergeometric function is
replaced by
\begin{equation}\label{}
    F\left(
    \frac{\ell}{2},\frac{\ell}{2}+1;\ell+\frac{3}{2};\bar{k}^2\right)
    \simeq \frac{2^{\ell+1/2}}{6\ell^2}.
\end{equation}
Furthermore, the Gamma functions in (\ref{eq:DeltaISWF}) can be
approximated using Stirling's formula~\cite{Abramovitz:1970aa}
\begin{equation}
\label{eq:stirling}
    \frac{\Gamma(\ell)}{\Gamma(\ell+3/2)} \simeq \frac{1}{\ell^{3/2}}.
\end{equation}
Putting everything together and squaring Eq. (\ref{eq:expAp}) we obtain
\begin{equation}\label{eq:Deltafinal}
\begin{aligned}
    |\Delta_\ell(\kb)|^2 & = 2\pi \Ao^2 \kb^{2n} \Om^{2n}
    \Bigg\{\left(1-\frac{\kb^2}{\kbe^2}\right)^2 \frac{\ue^{2\ve
    \sqrt{1-\kb^2}}}{\kb^3 (\vo-\ve)^3}
    \left(\besselJ{\ell+1/2}\left[\kb(\vo-\ve)\right]\right)^2
   \\  &+2
    \left(1-\frac{\kb^2}{\kbe^2}\right)
    \frac{\ue^{(\vo+\ve)\sqrt{1-\kb^2}}}{\kb^{3/2} (\vo-\ve)^{3/2}}
    \besselJ{\ell+1/2}\left[\kb(\vo-\ve)\right]
    \frac{\kb^{\ell-1}\sqrt{1-\kb^2}}{6\ell^{7/2}} \\  &+
    \ue^{2\vo\sqrt{1-\kb^2}}\frac{\kb^{2(\ell-1)}(1-\kb^2)}{36\ell^7}
    \Bigg \}
\end{aligned}
\end{equation}
The $\Cl$'s are then found by integrating over all k-modes
\begin{equation}\label{eq:Cldecomp}
\begin{aligned}
    \Cl &= \frac{2}{\pi}\ell(\ell+1) \Om^3 \int_0^1 \kb^2
    |\Delta_\ell(\kb)|^2 d \kb \\ & \equiv 4 \Ao^2
    \ell(\ell+1)\Om^{2n+3}\left(\Cl^{(1)} + \Cl^{(2)} + \Cl^{(3)}
    \right),
\end{aligned}
\end{equation}
where the $\Cl^{(i)}$ correspond to the three terms in
Eq.~(\ref{eq:Deltafinal}). In the following we keep only the zero
order terms in $\vo/\ve$. From Eqs.~(\ref{eq:Deltafinal}),
(\ref{eq:Cldecomp}) one finds
\begin{equation}\label{eq:Cl1}
    \Cl^{(1)} = \frac{1}{\vo^3}\int_0^1 \kb^{2n-1}
    \left(1-\frac{\kb^2}{\kbe^2}\right)^2 \ue^{2\ve \sqrt{1-\kb^2}}
    \left[\besselJ{\ell+1/2}\left(\kb \vo\right)\right]^2 \ud \kb
\end{equation}
First, notice that if the argument is larger or smaller than the
index, the Bessel functions are well approximated by their asymptotic
expansions (\ref{eq:bessellarge}) and (\ref{eq:besselsmall}),
respectively. Therefore, we split the $\kb$-integral into two
integrals over the intervals $[0,\kbl]$ and $[\kbl,1]$, in each of
which the Bessel function is replaced by its limiting forms. The
transition value $\kbl$ is given by $\kbl \simeq \ell/\vo$. In the
integral from $\kbl$ to $1$, the $\sin^2(\kb \vo)$ is then replaced by
its mean value $1/2$ which is justified if the multiplying function
varies much slower in $\kb$ than the sine.
To carry out the integration we make the substitution $y^2 =
1-\kb^2$, and in order to simplify the notation we define
the integral
\begin{equation}
\label{eq:defI}
    \Ic(a,b,\nu)=\int_a^b  y
    (1-y^2)^\nu e^{2\ve y}\ud y
\end{equation}
In this way we can write Eq.~(\ref{eq:Cl1}) in the form
\begin{equation}\label{eq:Cl1integrals}
\begin{aligned}
    \Cl^{(1)} = & \frac{1}{\pi \vo^4}
    \left[\Ic(0,\yl,n-3/2)-\frac{2}{\kbe^2} \Ic(0,\yl,n-1/2) +
\frac{1}{\kbe^4} \Ic(0,\yl,n+1/2)
    \right]\\
        & +\frac{1}{\Gamma^2(\ell+3/2)}\left(\dfrac{\vo}{2}\right)^{2\ell+1}
    \left[\Ic(\yl,1,\ell+n-1/2)-\frac{2}{\kbe^2} \Ic(\yl,1,\ell+n+1/2)
    + \frac{1}{\kbe^4} \Ic(\yl,1,\ell+n+3/2)
    \right]
\end{aligned}
\end{equation}
Since $\yl = \sqrt{1-\kbl^2}$ is very close to one, and the
integrand is continuous in the interval $[0,1]$, integrals of the
form $\Ic(\yl,1,\nu)$ can be  well approximated by the mean formula
\begin{equation}\label{eq:meanvalueintegral}
    \left. \Ic(\yl,1,\nu) \simeq y(1-y^2)^\nu e^{2\ve
    y}\right|_{y=\yl}(1-\yl)
    \simeq
    \frac{e^{2\ve}}{2}\left(\frac{\ell}{\vo}\right)^{2(\nu+1)}
\end{equation}
For the integrals of the type $\Ic(0,\yl,\nu)$ we distinguish between
three cases:
\begin{description}
\item[Case a: $\nu > -1$.] This case corresponds to a spectral index
$n> 1/2$ in the first integral in Eq.~(\ref{eq:Cl1}). We write
$\Ic(0,\yl,\nu) = \Ic(0,1,\nu)-\Ic(\yl,1,\nu)$. The solution of the
latter is given by Eq.~(\ref{eq:meanvalueintegral}), whereas the
former can be solved in terms of modified Bessel and Struve
functions~\cite{Gradshteyn:1965aa}
\begin{equation}\label{eq:Isol}
    \Ic(0,1,\nu) = \frac{1}{2(\nu+1)} +
    \frac{\sqrt{\pi}}{2}\ve^{-1/2-\nu}
    \Gamma(\nu+1)\left[\besselI{\nu+3/2}(2\ve)+\struveL{\nu+3/2}(2\ve)\right].
\end{equation}
Since our derivation assumes $\ve > \ell$, the large argument limit
applies and we have
\begin{equation}\label{}
    \besselI{\nu+3/2}(2\ve)+\struveL{\nu+3/2}(2\ve)
    \simeq \frac{e^{2\ve}}{\sqrt{\pi \ve}},
\end{equation}
independently of the index $\nu$.
\item[Case b: $\nu = -1$.]
Since  the above expressions, Eq.~(\ref{eq:Isol}), diverge for $\nu=-1$,
we approximate the integral by
\begin{equation}\label{eq:pulloutexpln}
    \Ic(0,\yl,\nu) \simeq e^{2\ve }\int_0^{\yl} y (1-y^2)^{-1} dy
    =-e^{2\ve} \ln\left(\frac{\ell}{\vo}\right)
\end{equation}
We have checked that the numerical solution of $\Ic(0,\yl,\nu)$ agrees
well with the approximation, provided $\yl$ is close to
$1$.
\item[Case c: $\nu < -1$.]
We use the same simplification as in Eq.~(\ref{eq:pulloutexpln}), and now
the integral yields
\begin{equation}\label{eq:pulloutexp}
    \Ic(0,\yl,\nu) \simeq e^{2\ve }\int_0^{\yl} y (1-y^2)^\nu \ud y
    =-\frac{e^{2\ve}}{2(\nu+1)} \left(\frac{\ell}{\vo}\right)^{2(\nu+1)}
\end{equation}
\end{description}
For the particular value $n=-3/2$, Eq.~(\ref{eq:Cl1integrals})
contains terms $\Ic(0,\yl,-3)$ and $\Ic(0,\yl,-2)$ which can be
evaluated according to (\ref{eq:pulloutexp}), as well as a term
$\Ic(0,\yl,-1)$ for which we use (\ref{eq:pulloutexpln}). The
remaining three integrals over the interval $[\yl,1]$ are evaluated by
(\ref{eq:meanvalueintegral}). The result is
\begin{equation}\label{eq:Cl1result}
    \Cl^{(1)} \simeq \dfrac{e^{\vo/\ze}}{4\pi\ell^4}\left[1-\frac{\ell^2}{6\ze}
    -\left(\frac{\ell^2}{12\ze}\right)^2 \ln\left(\frac{\ell}{\vo}\right)
    +\frac{e^{2 \beta \ell}}{2}\left(1-\frac{\ell^2}{24
    \ze}\right)^2\right]
\end{equation}
The parameter $\beta$ is a constant of order unity, within in our
approximation it  is
$\beta = 1-\ln2 \sim 0.3$.

The second term $\Cl^{(2)}$ in Eq.~(\ref{eq:Deltafinal}) reads
\begin{equation}
\begin{aligned}
\Cl^{(2)} & \simeq \dfrac{1}{3
\ell^{7/2}\vo^{3/2s}} \int_{0}^{1}
\ue^{(\vo+\ve)\sqrt{1-\kb^2}} \kb^{2n+\ell -1/2}
\sqrt{1- \kb^2} \left(1-\dfrac{\kb^2}{\kbe^2} \right) \besselJ{\ell + 1/2}
(\kb \vo) \ud \kb,
\end{aligned}
\end{equation}
where only the zero order terms in $\ve/\vo$ has been kept. Using the
limiting forms for the Bessel function for arguments smaller and larger
than the transition value $\kbl$, yields
\begin{equation}
\label{eq:cl2}
\begin{aligned}
\Cl^{(2)} & \simeq \dfrac{1}{3
\ell^{7/2}\vo^{3/2} \Gamma(\ell+3/2)} \int_{0}^{\kbl}
\ue^{(\vo+\ve)\sqrt{1-\kb^2}} \kb^{2n+\ell -1/2}
\sqrt{1- \kb^2} \left(1-\dfrac{\kb^2}{\kbe^2} \right)
\left(\dfrac{\kb \vo}{2} \right)^{\ell + 1/2} \ud \kb \\
& + \dfrac{2^{1/2}}{3 \pi^{1/2}
\ell^{7/2} \vo} \int_{\kbl}^{1} \ue^{(\vo+\ve)\sqrt{1-\kb^2}}
\kb^{2n+\ell -1} \sqrt{1- \kb^2} \left(1-\dfrac{\kb^2}{\kbe^2} \right)
\sin{\left(\kb \vo - \dfrac{\pi}{2} \ell \right)} \ud \kb.
\end{aligned}
\end{equation}
For consistency with the derivation of $\Cl^{(1)}$, we have assumed
that the main contribution comes from the first integral, while the
second one is small due to the oscillating integrand. Since $\kbl \ll
1$, we can use again the mean formula to evaluate the first integral,
and by the Stirling formula for $\Gamma(\ell+3/2)$, Eq.~(\ref{eq:cl2})
becomes
\begin{equation}
\label{eq:Cl2result}
\Cl^{(2)} \simeq \dfrac{\ue^{\vo}}{12
\pi^{1/2}  \ell^{11/2}} \left(\dfrac{\ell}{\vo}\right)^{2n + \ell +2}
\ue^{\beta \ell} \left(1 - \dfrac{\ell^2}{24 \ze} \right).
\end{equation}
Since $\ell < \vo$, the spectrum is damped at large $\ell$, while the
other terms can lead to the appearance of a bump, depending on the value
of $\vo$ and $n$.

The last terms $\Cl^{(3)}$ reads
\begin{equation}
\begin{aligned}
\Cl^{(3)} & = \dfrac{1}{36 \ell^7 } \int_{0}^{1}
\ue^{2\vo\sqrt{1-\kb^2}} \kb^{2n + 2\ell}
\left(1- \kb^2 \right) \ud \kb.
\end{aligned}
\end{equation}
Splitting this expression in two terms over $1-\kb^2$, and using the
substitution $y^2=1-\kb^2$ yields
\begin{equation}
\Cl^{(3)} =  \dfrac{1}{36 \ell^7} \left[\Ic\left(0,1,n+\ell-\dfrac{1}{2}
\right) - \Ic\left(0,1,n + \ell + \dfrac{1}{2} \right) \right],
\end{equation}
where $\Ic$ is given by Eq.~(\ref{eq:defI}) with $\ve \rightarrow
\vo$. As before, these two integrals can be expressed in terms of
modified Bessel and Struve functions~\cite{Gradshteyn:1965aa}. From
Eq.~(\ref{eq:Isol}), taking their limiting forms at large argument,
and expanding the $\Gamma$ function by means of the Stirling formula
gives
\begin{equation}
\label{eq:Cl3result}
\Cl^{(3)} \simeq \dfrac{1}{72 \ell^7 \left(n+\ell+3/2\right)
\left(n +\ell + 1/2 \right)} + \sqrt{\dfrac{\pi}{2}} 
\dfrac{\ue^{2 \vo}}{\ell^{15/2}} \dfrac{\ue^{-\ell}}{36}
\left(\dfrac{\ell}{\vo}\right)^{n + \ell + 1/2}.
\end{equation}
Clearly, $\Cl^{(3)}$ dominates over the others since it involves
$\exp(2\vo)$ while $\Cl^{(2)}$ and $\Cl^{(1)}$ appear only with
fractional power of this factor, namely $\exp(\vo)$ and
$\exp(\vo/\ze)$. This is due to the fact that we are concerned with
incessantly growing perturbations leading to the predominance of the
integrated Sachs-Wolfe effect.

Inserting Eqs.~(\ref{eq:Cl1result}), (\ref{eq:Cl2result}) and
(\ref{eq:Cl3result}) for the particular value $n=-3/2$ into
Eq.~(\ref{eq:Cldecomp}) gives the final CMB angular power spectrum
\begin{equation}
\begin{aligned}
\dfrac{\ell (\ell+1)}{2\pi} \Cl &\simeq \dfrac{2}{\pi} \Ao^2
\Bigg\{\dfrac{e^{\vo/\ze}}{4\pi}\left[1-\frac{\ell^2}{6\ze}
    -\left(\frac{\ell^2}{12\ze}\right)^2 \ln\left(\frac{\ell}{\vo}\right)
    +\frac{e^{2 \beta \ell}}{2}\left(1-\frac{\ell^2}{24
    \ze}\right)^2\right] \\
& + \dfrac{\ue^{\vo}}{12
\pi^{1/2} \ell^{3/2}} \left(\dfrac{\ell}{\vo}\right)^{\ell-1} \ue^{\beta
\ell} \left(1 - \dfrac{\ell^2}{24 \ze} \right)
 + \sqrt{\dfrac{\pi}{2}} 
\dfrac{\ue^{2 \vo}}{\ell^{7/2}} \dfrac{\ue^{-\ell}}{36}
\left(\dfrac{\ell}{\vo}\right)^{\ell-1} \Bigg\}.
\end{aligned}
\end{equation}

\end{widetext}

\bibliography{bibfile}

\begin{thebibliography}{84}
\expandafter\ifx\csname natexlab\endcsname\relax\def\natexlab#1{#1}\fi
\expandafter\ifx\csname bibnamefont\endcsname\relax
  \def\bibnamefont#1{#1}\fi
\expandafter\ifx\csname bibfnamefont\endcsname\relax
  \def\bibfnamefont#1{#1}\fi
\expandafter\ifx\csname citenamefont\endcsname\relax
  \def\citenamefont#1{#1}\fi
\expandafter\ifx\csname url\endcsname\relax
  \def\url#1{\texttt{#1}}\fi
\expandafter\ifx\csname urlprefix\endcsname\relax\def\urlprefix{URL }\fi
\providecommand{\bibinfo}[2]{#2}
\providecommand{\eprint}[2][]{\url{#2}}

\bibitem[{\citenamefont{Nordstr{\"o}m}(1914)}]{Nordstrom:1914}
\bibinfo{author}{\bibfnamefont{G.}~\bibnamefont{Nordstr{\"o}m}},
  \bibinfo{journal}{Phys. Zeitschr.} \textbf{\bibinfo{volume}{15}},
  \bibinfo{pages}{504} (\bibinfo{year}{1914}).

\bibitem[{\citenamefont{Kaluza}(1921)}]{Kaluza:1921}
\bibinfo{author}{\bibfnamefont{T.}~\bibnamefont{Kaluza}},
  \bibinfo{journal}{Sitzungsber. Preuss. Akad. Wiss. Berlin} p.
  \bibinfo{pages}{966} (\bibinfo{year}{1921}).

\bibitem[{\citenamefont{Klein}(1926)}]{Klein:1926}
\bibinfo{author}{\bibfnamefont{O.}~\bibnamefont{Klein}}, \bibinfo{journal}{Z.
  Phys.} \textbf{\bibinfo{volume}{37}}, \bibinfo{pages}{895}
  (\bibinfo{year}{1926}).

\bibitem[{\citenamefont{Polchinski}(1998{\natexlab{a}})}]{Polchinski:1998rq}
\bibinfo{author}{\bibfnamefont{J.}~\bibnamefont{Polchinski}},
  \emph{\bibinfo{title}{String theory. An introduction to the bosonic string,
  Vol. I}} (\bibinfo{publisher}{Cambridge University Press},
  \bibinfo{address}{Cambridge, UK}, \bibinfo{year}{1998}{\natexlab{a}}).

\bibitem[{\citenamefont{Polchinski}(1998{\natexlab{b}})}]{Polchinski:1998rr}
\bibinfo{author}{\bibfnamefont{J.}~\bibnamefont{Polchinski}},
  \emph{\bibinfo{title}{String theory. Superstring theory and beyond, Vol. II}}
  (\bibinfo{publisher}{Cambridge University Press},
  \bibinfo{address}{Cambridge, UK}, \bibinfo{year}{1998}{\natexlab{b}}).

\bibitem[{\citenamefont{Horava and Witten}(1996)}]{Horava:1996qa}
\bibinfo{author}{\bibfnamefont{P.}~\bibnamefont{Horava}} \bibnamefont{and}
  \bibinfo{author}{\bibfnamefont{E.}~\bibnamefont{Witten}},
  \bibinfo{journal}{Nucl. Phys.} \textbf{\bibinfo{volume}{B460}},
  \bibinfo{pages}{506} (\bibinfo{year}{1996}),
  \eprint[http://arXiv.org/abs]{hep-th/9510209}.

\bibitem[{\citenamefont{Antoniadis}(1990)}]{Antoniadis:1990ew}
\bibinfo{author}{\bibfnamefont{I.}~\bibnamefont{Antoniadis}},
  \bibinfo{journal}{Phys. Lett.} \textbf{\bibinfo{volume}{B246}},
  \bibinfo{pages}{377} (\bibinfo{year}{1990}).

\bibitem[{\citenamefont{Polchinski}(1995)}]{Polchinski:1995mt}
\bibinfo{author}{\bibfnamefont{J.}~\bibnamefont{Polchinski}},
  \bibinfo{journal}{Phys. Rev. Lett.} \textbf{\bibinfo{volume}{75}},
  \bibinfo{pages}{4724} (\bibinfo{year}{1995}), \eprint{hep-th/9510017}.

\bibitem[{\citenamefont{Lukas et~al.}(1999)\citenamefont{Lukas, Ovrut, and
  Waldram}}]{Lukas:1998qs}
\bibinfo{author}{\bibfnamefont{A.}~\bibnamefont{Lukas}},
  \bibinfo{author}{\bibfnamefont{B.~A.} \bibnamefont{Ovrut}}, \bibnamefont{and}
  \bibinfo{author}{\bibfnamefont{D.}~\bibnamefont{Waldram}},
  \bibinfo{journal}{Phys. Rev.} \textbf{\bibinfo{volume}{D60}},
  \bibinfo{pages}{086001} (\bibinfo{year}{1999}), \eprint{hep-th/9806022}.

\bibitem[{\citenamefont{Akama}(1982)}]{Akama:1982jy}
\bibinfo{author}{\bibfnamefont{K.}~\bibnamefont{Akama}},
  \bibinfo{journal}{Lect. Notes Phys.} \textbf{\bibinfo{volume}{176}},
  \bibinfo{pages}{267} (\bibinfo{year}{1982}), \eprint{hep-th/0001113}.

\bibitem[{\citenamefont{Rubakov and Shaposhnikov}(1983)}]{Rubakov:1983bb}
\bibinfo{author}{\bibfnamefont{V.~A.} \bibnamefont{Rubakov}} \bibnamefont{and}
  \bibinfo{author}{\bibfnamefont{M.~E.} \bibnamefont{Shaposhnikov}},
  \bibinfo{journal}{Phys. Lett.} \textbf{\bibinfo{volume}{B125}},
  \bibinfo{pages}{136} (\bibinfo{year}{1983}).

\bibitem[{\citenamefont{Arkani-Hamed et~al.}(1998)\citenamefont{Arkani-Hamed,
  Dimopoulos, and Dvali}}]{Arkani-Hamed:1998rs}
\bibinfo{author}{\bibfnamefont{N.}~\bibnamefont{Arkani-Hamed}},
  \bibinfo{author}{\bibfnamefont{S.}~\bibnamefont{Dimopoulos}},
  \bibnamefont{and} \bibinfo{author}{\bibfnamefont{G.~R.} \bibnamefont{Dvali}},
  \bibinfo{journal}{Phys. Lett.} \textbf{\bibinfo{volume}{B429}},
  \bibinfo{pages}{263} (\bibinfo{year}{1998}), \eprint{hep-ph/9803315}.

\bibitem[{\citenamefont{Arkani-Hamed et~al.}(1999)\citenamefont{Arkani-Hamed,
  Dimopoulos, and Dvali}}]{Arkani-Hamed:1998nn}
\bibinfo{author}{\bibfnamefont{N.}~\bibnamefont{Arkani-Hamed}},
  \bibinfo{author}{\bibfnamefont{S.}~\bibnamefont{Dimopoulos}},
  \bibnamefont{and} \bibinfo{author}{\bibfnamefont{G.~R.} \bibnamefont{Dvali}},
  \bibinfo{journal}{Phys. Rev.} \textbf{\bibinfo{volume}{D59}},
  \bibinfo{pages}{086004} (\bibinfo{year}{1999}), \eprint{hep-ph/9807344}.

\bibitem[{\citenamefont{Antoniadis et~al.}(1998)\citenamefont{Antoniadis,
  Arkani-Hamed, Dimopoulos, and Dvali}}]{Antoniadis:1998ig}
\bibinfo{author}{\bibfnamefont{I.}~\bibnamefont{Antoniadis}},
  \bibinfo{author}{\bibfnamefont{N.}~\bibnamefont{Arkani-Hamed}},
  \bibinfo{author}{\bibfnamefont{S.}~\bibnamefont{Dimopoulos}},
  \bibnamefont{and} \bibinfo{author}{\bibfnamefont{G.~R.} \bibnamefont{Dvali}},
  \bibinfo{journal}{Phys. Lett.} \textbf{\bibinfo{volume}{B436}},
  \bibinfo{pages}{257} (\bibinfo{year}{1998}), \eprint{hep-ph/9804398}.

\bibitem[{\citenamefont{Randall and
  Sundrum}(1999{\natexlab{a}})}]{Randall:1999ee}
\bibinfo{author}{\bibfnamefont{L.}~\bibnamefont{Randall}} \bibnamefont{and}
  \bibinfo{author}{\bibfnamefont{R.}~\bibnamefont{Sundrum}},
  \bibinfo{journal}{Phys. Rev. Lett.} \textbf{\bibinfo{volume}{83}},
  \bibinfo{pages}{3370} (\bibinfo{year}{1999}{\natexlab{a}}),
  \eprint{hep-ph/9905221}.

\bibitem[{\citenamefont{Randall and
  Sundrum}(1999{\natexlab{b}})}]{Randall:1999vf}
\bibinfo{author}{\bibfnamefont{L.}~\bibnamefont{Randall}} \bibnamefont{and}
  \bibinfo{author}{\bibfnamefont{R.}~\bibnamefont{Sundrum}},
  \bibinfo{journal}{Phys. Rev. Lett.} \textbf{\bibinfo{volume}{83}},
  \bibinfo{pages}{4690} (\bibinfo{year}{1999}{\natexlab{b}}),
  \eprint{hep-th/9906064}.

\bibitem[{\citenamefont{Long et~al.}(2002)}]{Long:2002wn}
\bibinfo{author}{\bibfnamefont{J.~C.} \bibnamefont{Long}} \bibnamefont{et~al.}
  (\bibinfo{year}{2002}), \eprint{hep-ph/0210004}.

\bibitem[{\citenamefont{Uzan and Bernardeau}(2001)}]{Uzan:2000mz}
\bibinfo{author}{\bibfnamefont{J.-P.} \bibnamefont{Uzan}} \bibnamefont{and}
  \bibinfo{author}{\bibfnamefont{F.}~\bibnamefont{Bernardeau}},
  \bibinfo{journal}{Phys. Rev.} \textbf{\bibinfo{volume}{D64}},
  \bibinfo{pages}{083004} (\bibinfo{year}{2001}), \eprint{hep-ph/0012011}.

\bibitem[{\citenamefont{Allen et~al.}(2000)\citenamefont{Allen, Ettori, and
  Fabian}}]{Allen:2000ih}
\bibinfo{author}{\bibfnamefont{S.~W.} \bibnamefont{Allen}},
  \bibinfo{author}{\bibfnamefont{S.}~\bibnamefont{Ettori}}, \bibnamefont{and}
  \bibinfo{author}{\bibfnamefont{A.~C.} \bibnamefont{Fabian}}
  (\bibinfo{year}{2000}), \eprint{astro-ph/0008517}.

\bibitem[{\citenamefont{Binetruy
  et~al.}(2000{\natexlab{a}})\citenamefont{Binetruy, Deffayet, and
  Langlois}}]{Binetruy:1999ut}
\bibinfo{author}{\bibfnamefont{P.}~\bibnamefont{Binetruy}},
  \bibinfo{author}{\bibfnamefont{C.}~\bibnamefont{Deffayet}}, \bibnamefont{and}
  \bibinfo{author}{\bibfnamefont{D.}~\bibnamefont{Langlois}},
  \bibinfo{journal}{Nucl. Phys.} \textbf{\bibinfo{volume}{B565}},
  \bibinfo{pages}{269} (\bibinfo{year}{2000}{\natexlab{a}}),
  \eprint[http://arXiv.org/abs]{hep-th/9905012}.

\bibitem[{\citenamefont{Csaki et~al.}(1999)\citenamefont{Csaki, Graesser,
  Kolda, and Terning}}]{Csaki:1999jh}
\bibinfo{author}{\bibfnamefont{C.}~\bibnamefont{Csaki}},
  \bibinfo{author}{\bibfnamefont{M.}~\bibnamefont{Graesser}},
  \bibinfo{author}{\bibfnamefont{C.~F.} \bibnamefont{Kolda}}, \bibnamefont{and}
  \bibinfo{author}{\bibfnamefont{J.}~\bibnamefont{Terning}},
  \bibinfo{journal}{Phys. Lett.} \textbf{\bibinfo{volume}{B462}},
  \bibinfo{pages}{34} (\bibinfo{year}{1999}),
  \eprint[http://arXiv.org/abs]{hep-ph/9906513}.

\bibitem[{\citenamefont{Cline et~al.}(1999)\citenamefont{Cline, Grojean, and
  Servant}}]{Cline:1999ts}
\bibinfo{author}{\bibfnamefont{J.~M.} \bibnamefont{Cline}},
  \bibinfo{author}{\bibfnamefont{C.}~\bibnamefont{Grojean}}, \bibnamefont{and}
  \bibinfo{author}{\bibfnamefont{G.}~\bibnamefont{Servant}},
  \bibinfo{journal}{Phys. Rev. Lett.} \textbf{\bibinfo{volume}{83}},
  \bibinfo{pages}{4245} (\bibinfo{year}{1999}),
  \eprint[http://arXiv.org/abs]{hep-ph/9906523}.

\bibitem[{\citenamefont{Shiromizu et~al.}(2000)\citenamefont{Shiromizu, Maeda,
  and Sasaki}}]{Shiromizu:1999wj}
\bibinfo{author}{\bibfnamefont{T.}~\bibnamefont{Shiromizu}},
  \bibinfo{author}{\bibfnamefont{K.-i.} \bibnamefont{Maeda}}, \bibnamefont{and}
  \bibinfo{author}{\bibfnamefont{M.}~\bibnamefont{Sasaki}},
  \bibinfo{journal}{Phys. Rev.} \textbf{\bibinfo{volume}{D62}},
  \bibinfo{pages}{024012} (\bibinfo{year}{2000}),
  \eprint[http://arXiv.org/abs]{gr-qc/9910076}.

\bibitem[{\citenamefont{Flanagan et~al.}(2000)\citenamefont{Flanagan, Tye, and
  Wasserman}}]{Flanagan:1999cu}
\bibinfo{author}{\bibfnamefont{E.~E.} \bibnamefont{Flanagan}},
  \bibinfo{author}{\bibfnamefont{S.~H.~H.} \bibnamefont{Tye}},
  \bibnamefont{and}
  \bibinfo{author}{\bibfnamefont{I.}~\bibnamefont{Wasserman}},
  \bibinfo{journal}{Phys. Rev.} \textbf{\bibinfo{volume}{D62}},
  \bibinfo{pages}{044039} (\bibinfo{year}{2000}), \eprint{hep-ph/9910498}.

\bibitem[{\citenamefont{Maartens et~al.}(2000)\citenamefont{Maartens, Wands,
  Bassett, and Heard}}]{Maartens:1999hf}
\bibinfo{author}{\bibfnamefont{R.}~\bibnamefont{Maartens}},
  \bibinfo{author}{\bibfnamefont{D.}~\bibnamefont{Wands}},
  \bibinfo{author}{\bibfnamefont{B.~A.} \bibnamefont{Bassett}},
  \bibnamefont{and} \bibinfo{author}{\bibfnamefont{I.}~\bibnamefont{Heard}},
  \bibinfo{journal}{Phys. Rev.} \textbf{\bibinfo{volume}{D62}},
  \bibinfo{pages}{041301} (\bibinfo{year}{2000}), \eprint{hep-ph/9912464}.

\bibitem[{\citenamefont{Rubakov}(2001)}]{Rubakov:2001kp}
\bibinfo{author}{\bibfnamefont{V.~A.} \bibnamefont{Rubakov}},
  \bibinfo{journal}{Phys. Usp.} \textbf{\bibinfo{volume}{44}},
  \bibinfo{pages}{871} (\bibinfo{year}{2001}), \eprint{hep-ph/0104152}.

\bibitem[{\citenamefont{Carter and Uzan}(2001)}]{Carter:2001nj}
\bibinfo{author}{\bibfnamefont{B.}~\bibnamefont{Carter}} \bibnamefont{and}
  \bibinfo{author}{\bibfnamefont{J.-P.} \bibnamefont{Uzan}},
  \bibinfo{journal}{Nucl. Phys.} \textbf{\bibinfo{volume}{B606}},
  \bibinfo{pages}{45} (\bibinfo{year}{2001}), \eprint{gr-qc/0101010}.

\bibitem[{\citenamefont{Battye et~al.}(2001)\citenamefont{Battye, Carter,
  Mennim, and Uzan}}]{Battye:2001yn}
\bibinfo{author}{\bibfnamefont{R.~A.} \bibnamefont{Battye}},
  \bibinfo{author}{\bibfnamefont{B.}~\bibnamefont{Carter}},
  \bibinfo{author}{\bibfnamefont{A.}~\bibnamefont{Mennim}}, \bibnamefont{and}
  \bibinfo{author}{\bibfnamefont{J.-P.} \bibnamefont{Uzan}},
  \bibinfo{journal}{Phys. Rev.} \textbf{\bibinfo{volume}{D64}},
  \bibinfo{pages}{124007} (\bibinfo{year}{2001}), \eprint{hep-th/0105091}.

\bibitem[{\citenamefont{Carter et~al.}(2001)\citenamefont{Carter, Uzan, Battye,
  and Mennim}}]{Carter:2001af}
\bibinfo{author}{\bibfnamefont{B.}~\bibnamefont{Carter}},
  \bibinfo{author}{\bibfnamefont{J.-P.} \bibnamefont{Uzan}},
  \bibinfo{author}{\bibfnamefont{R.~A.} \bibnamefont{Battye}},
  \bibnamefont{and} \bibinfo{author}{\bibfnamefont{A.}~\bibnamefont{Mennim}},
  \bibinfo{journal}{Class. Quant. Grav.} \textbf{\bibinfo{volume}{18}},
  \bibinfo{pages}{4871} (\bibinfo{year}{2001}),
  \eprint[http://arXiv.org/abs]{gr-qc/0106038}.

\bibitem[{\citenamefont{Binetruy
  et~al.}(2000{\natexlab{b}})\citenamefont{Binetruy, Deffayet, Ellwanger, and
  Langlois}}]{Binetruy:1999hy}
\bibinfo{author}{\bibfnamefont{P.}~\bibnamefont{Binetruy}},
  \bibinfo{author}{\bibfnamefont{C.}~\bibnamefont{Deffayet}},
  \bibinfo{author}{\bibfnamefont{U.}~\bibnamefont{Ellwanger}},
  \bibnamefont{and} \bibinfo{author}{\bibfnamefont{D.}~\bibnamefont{Langlois}},
  \bibinfo{journal}{Phys. Lett.} \textbf{\bibinfo{volume}{B477}},
  \bibinfo{pages}{285} (\bibinfo{year}{2000}{\natexlab{b}}),
  \eprint{hep-th/9910219}.

\bibitem[{\citenamefont{Vollick}(2001)}]{Vollick:1999uz}
\bibinfo{author}{\bibfnamefont{D.~N.} \bibnamefont{Vollick}},
  \bibinfo{journal}{Class. Quant. Grav.} \textbf{\bibinfo{volume}{18}},
  \bibinfo{pages}{1} (\bibinfo{year}{2001}), \eprint{hep-th/9911181}.

\bibitem[{\citenamefont{Kraus}(1999)}]{Kraus:1999it}
\bibinfo{author}{\bibfnamefont{P.}~\bibnamefont{Kraus}},
  \bibinfo{journal}{JHEP} \textbf{\bibinfo{volume}{12}}, \bibinfo{pages}{011}
  (\bibinfo{year}{1999}), \eprint[http://arXiv.org/abs]{hep-th/9910149}.

\bibitem[{\citenamefont{Ida}(2000)}]{Ida:1999ui}
\bibinfo{author}{\bibfnamefont{D.}~\bibnamefont{Ida}}, \bibinfo{journal}{JHEP}
  \textbf{\bibinfo{volume}{09}}, \bibinfo{pages}{014} (\bibinfo{year}{2000}),
  \eprint{gr-qc/9912002}.

\bibitem[{\citenamefont{Mukohyama et~al.}(2000)\citenamefont{Mukohyama,
  Shiromizu, and Maeda}}]{Mukohyama:1999wi}
\bibinfo{author}{\bibfnamefont{S.}~\bibnamefont{Mukohyama}},
  \bibinfo{author}{\bibfnamefont{T.}~\bibnamefont{Shiromizu}},
  \bibnamefont{and} \bibinfo{author}{\bibfnamefont{K.-i.} \bibnamefont{Maeda}},
  \bibinfo{journal}{Phys. Rev.} \textbf{\bibinfo{volume}{D62}},
  \bibinfo{pages}{024028} (\bibinfo{year}{2000}), \eprint{hep-th/9912287}.

\bibitem[{\citenamefont{Bonjour et~al.}(1999)\citenamefont{Bonjour, Charmousis,
  and Gregory}}]{Bonjour:1999kz}
\bibinfo{author}{\bibfnamefont{F.}~\bibnamefont{Bonjour}},
  \bibinfo{author}{\bibfnamefont{C.}~\bibnamefont{Charmousis}},
  \bibnamefont{and} \bibinfo{author}{\bibfnamefont{R.}~\bibnamefont{Gregory}},
  \bibinfo{journal}{Class. Quant. Grav.} \textbf{\bibinfo{volume}{16}},
  \bibinfo{pages}{2427} (\bibinfo{year}{1999}), \eprint{gr-qc/9902081}.

\bibitem[{\citenamefont{Gregory}(2000)}]{Gregory:1999gv}
\bibinfo{author}{\bibfnamefont{R.}~\bibnamefont{Gregory}},
  \bibinfo{journal}{Phys. Rev. Lett.} \textbf{\bibinfo{volume}{84}},
  \bibinfo{pages}{2564} (\bibinfo{year}{2000}), \eprint{hep-th/9911015}.

\bibitem[{\citenamefont{Antunes et~al.}(2002)\citenamefont{Antunes, Copeland,
  Hindmarsh, and Lukas}}]{Antunes:2002hn}
\bibinfo{author}{\bibfnamefont{N.~D.} \bibnamefont{Antunes}},
  \bibinfo{author}{\bibfnamefont{E.~J.} \bibnamefont{Copeland}},
  \bibinfo{author}{\bibfnamefont{M.}~\bibnamefont{Hindmarsh}},
  \bibnamefont{and} \bibinfo{author}{\bibfnamefont{A.}~\bibnamefont{Lukas}}
  (\bibinfo{year}{2002}), \eprint{hep-th/0208219}.

\bibitem[{\citenamefont{Nihei}(2000)}]{Nihei:2000gb}
\bibinfo{author}{\bibfnamefont{T.}~\bibnamefont{Nihei}},
  \bibinfo{journal}{Phys. Rev.} \textbf{\bibinfo{volume}{D62}},
  \bibinfo{pages}{124017} (\bibinfo{year}{2000}), \eprint{hep-th/0005014}.

\bibitem[{\citenamefont{Gherghetta and Shaposhnikov}(2000)}]{Gherghetta:2000qi}
\bibinfo{author}{\bibfnamefont{T.}~\bibnamefont{Gherghetta}} \bibnamefont{and}
  \bibinfo{author}{\bibfnamefont{M.~E.} \bibnamefont{Shaposhnikov}},
  \bibinfo{journal}{Phys. Rev. Lett.} \textbf{\bibinfo{volume}{85}},
  \bibinfo{pages}{240} (\bibinfo{year}{2000}), \eprint{hep-th/0004014}.

\bibitem[{\citenamefont{Ringeval et~al.}(2002)\citenamefont{Ringeval, Peter,
  and Uzan}}]{Ringeval:2001cq}
\bibinfo{author}{\bibfnamefont{C.}~\bibnamefont{Ringeval}},
  \bibinfo{author}{\bibfnamefont{P.}~\bibnamefont{Peter}}, \bibnamefont{and}
  \bibinfo{author}{\bibfnamefont{J.-P.} \bibnamefont{Uzan}},
  \bibinfo{journal}{Phys. Rev.} \textbf{\bibinfo{volume}{D65}},
  \bibinfo{pages}{044016} (\bibinfo{year}{2002}), \eprint{hep-th/0109194}.

\bibitem[{\citenamefont{Ringeval}(2001)}]{Ringeval:2001xd}
\bibinfo{author}{\bibfnamefont{C.}~\bibnamefont{Ringeval}},
  \bibinfo{journal}{Phys. Rev.} \textbf{\bibinfo{volume}{D64}},
  \bibinfo{pages}{123505} (\bibinfo{year}{2001}), \eprint{hep-ph/0106179}.

\bibitem[{\citenamefont{Bajc and Gabadadze}(2000)}]{Bajc:1999mh}
\bibinfo{author}{\bibfnamefont{B.}~\bibnamefont{Bajc}} \bibnamefont{and}
  \bibinfo{author}{\bibfnamefont{G.}~\bibnamefont{Gabadadze}},
  \bibinfo{journal}{Phys. Lett.} \textbf{\bibinfo{volume}{B474}},
  \bibinfo{pages}{282} (\bibinfo{year}{2000}), \eprint{hep-th/9912232}.

\bibitem[{\citenamefont{Dvali and Shifman}(1997)}]{Dvali:1997xe}
\bibinfo{author}{\bibfnamefont{G.~R.} \bibnamefont{Dvali}} \bibnamefont{and}
  \bibinfo{author}{\bibfnamefont{M.~A.} \bibnamefont{Shifman}},
  \bibinfo{journal}{Phys. Lett.} \textbf{\bibinfo{volume}{B396}},
  \bibinfo{pages}{64} (\bibinfo{year}{1997}), \eprint{hep-th/9612128}.

\bibitem[{\citenamefont{Dubovsky et~al.}(2000)\citenamefont{Dubovsky, Rubakov,
  and Tinyakov}}]{Dubovsky:2000am}
\bibinfo{author}{\bibfnamefont{S.~L.} \bibnamefont{Dubovsky}},
  \bibinfo{author}{\bibfnamefont{V.~A.} \bibnamefont{Rubakov}},
  \bibnamefont{and} \bibinfo{author}{\bibfnamefont{P.~G.}
  \bibnamefont{Tinyakov}}, \bibinfo{journal}{Phys. Rev.}
  \textbf{\bibinfo{volume}{D62}}, \bibinfo{pages}{105011}
  (\bibinfo{year}{2000}), \eprint{hep-th/0006046}.

\bibitem[{\citenamefont{Dvali et~al.}(2001)\citenamefont{Dvali, Gabadadze, and
  Shifman}}]{Dvali:2000rx}
\bibinfo{author}{\bibfnamefont{G.~R.} \bibnamefont{Dvali}},
  \bibinfo{author}{\bibfnamefont{G.}~\bibnamefont{Gabadadze}},
  \bibnamefont{and} \bibinfo{author}{\bibfnamefont{M.~A.}
  \bibnamefont{Shifman}}, \bibinfo{journal}{Phys. Lett.}
  \textbf{\bibinfo{volume}{B497}}, \bibinfo{pages}{271} (\bibinfo{year}{2001}),
  \eprint{hep-th/0010071}.

\bibitem[{\citenamefont{Dimopoulos et~al.}(2001)\citenamefont{Dimopoulos,
  Farakos, Kehagias, and Koutsoumbas}}]{Dimopoulos:2000ej}
\bibinfo{author}{\bibfnamefont{P.}~\bibnamefont{Dimopoulos}},
  \bibinfo{author}{\bibfnamefont{K.}~\bibnamefont{Farakos}},
  \bibinfo{author}{\bibfnamefont{A.}~\bibnamefont{Kehagias}}, \bibnamefont{and}
  \bibinfo{author}{\bibfnamefont{G.}~\bibnamefont{Koutsoumbas}},
  \bibinfo{journal}{Nucl. Phys.} \textbf{\bibinfo{volume}{B617}},
  \bibinfo{pages}{237} (\bibinfo{year}{2001}), \eprint{hep-th/0007079}.

\bibitem[{\citenamefont{Duff et~al.}(2001)\citenamefont{Duff, Liu, and
  Sabra}}]{Duff:2000jk}
\bibinfo{author}{\bibfnamefont{M.~J.} \bibnamefont{Duff}},
  \bibinfo{author}{\bibfnamefont{J.~T.} \bibnamefont{Liu}}, \bibnamefont{and}
  \bibinfo{author}{\bibfnamefont{W.~A.} \bibnamefont{Sabra}},
  \bibinfo{journal}{Nucl. Phys.} \textbf{\bibinfo{volume}{B605}},
  \bibinfo{pages}{234} (\bibinfo{year}{2001}), \eprint{hep-th/0009212}.

\bibitem[{\citenamefont{Oda}(2001)}]{Oda:2001ux}
\bibinfo{author}{\bibfnamefont{I.}~\bibnamefont{Oda}} (\bibinfo{year}{2001}),
  \eprint{hep-th/0103052}.

\bibitem[{\citenamefont{Ghoroku and Nakamura}(2002)}]{Ghoroku:2001zu}
\bibinfo{author}{\bibfnamefont{K.}~\bibnamefont{Ghoroku}} \bibnamefont{and}
  \bibinfo{author}{\bibfnamefont{A.}~\bibnamefont{Nakamura}},
  \bibinfo{journal}{Phys. Rev.} \textbf{\bibinfo{volume}{D65}},
  \bibinfo{pages}{084017} (\bibinfo{year}{2002}), \eprint{hep-th/0106145}.

\bibitem[{\citenamefont{Akhmedov}(2001)}]{Akhmedov:2001ny}
\bibinfo{author}{\bibfnamefont{E.~K.} \bibnamefont{Akhmedov}},
  \bibinfo{journal}{Phys. Lett.} \textbf{\bibinfo{volume}{B521}},
  \bibinfo{pages}{79} (\bibinfo{year}{2001}), \eprint{hep-th/0107223}.

\bibitem[{\citenamefont{Neronov}(2002)}]{Neronov:2001qv}
\bibinfo{author}{\bibfnamefont{A.}~\bibnamefont{Neronov}},
  \bibinfo{journal}{Phys. Rev.} \textbf{\bibinfo{volume}{D65}},
  \bibinfo{pages}{044004} (\bibinfo{year}{2002}), \eprint{gr-qc/0106092}.

\bibitem[{\citenamefont{Gherghetta et~al.}(2000)\citenamefont{Gherghetta,
  Roessl, and Shaposhnikov}}]{Gherghetta:2000jf}
\bibinfo{author}{\bibfnamefont{T.}~\bibnamefont{Gherghetta}},
  \bibinfo{author}{\bibfnamefont{E.}~\bibnamefont{Roessl}}, \bibnamefont{and}
  \bibinfo{author}{\bibfnamefont{M.~E.} \bibnamefont{Shaposhnikov}},
  \bibinfo{journal}{Phys. Lett.} \textbf{\bibinfo{volume}{B491}},
  \bibinfo{pages}{353} (\bibinfo{year}{2000}), \eprint{hep-th/0006251}.

\bibitem[{\citenamefont{Kanti et~al.}(2001)\citenamefont{Kanti, Madden, and
  Olive}}]{Kanti:2001vb}
\bibinfo{author}{\bibfnamefont{P.}~\bibnamefont{Kanti}},
  \bibinfo{author}{\bibfnamefont{R.}~\bibnamefont{Madden}}, \bibnamefont{and}
  \bibinfo{author}{\bibfnamefont{K.~A.} \bibnamefont{Olive}},
  \bibinfo{journal}{Phys. Rev.} \textbf{\bibinfo{volume}{D64}},
  \bibinfo{pages}{044021} (\bibinfo{year}{2001}), \eprint{hep-th/0104177}.

\bibitem[{\citenamefont{Peter et~al.}(2003)\citenamefont{Peter, Ringeval, and
  Uzan}}]{Peter:2003zg}
\bibinfo{author}{\bibfnamefont{P.}~\bibnamefont{Peter}},
  \bibinfo{author}{\bibfnamefont{C.}~\bibnamefont{Ringeval}}, \bibnamefont{and}
  \bibinfo{author}{\bibfnamefont{J.-P.} \bibnamefont{Uzan}}
  (\bibinfo{year}{2003}), \eprint{hep-th/0301172}.

\bibitem[{\citenamefont{Giovannini}(2001)}]{Giovannini:2001fh}
\bibinfo{author}{\bibfnamefont{M.}~\bibnamefont{Giovannini}},
  \bibinfo{journal}{Phys. Rev.} \textbf{\bibinfo{volume}{D64}},
  \bibinfo{pages}{064023} (\bibinfo{year}{2001}), \eprint{hep-th/0106041}.

\bibitem[{\citenamefont{Boehm et~al.}(2001)\citenamefont{Boehm, Durrer, and
  van~de Bruck}}]{Boehm:2001sp}
\bibinfo{author}{\bibfnamefont{T.}~\bibnamefont{Boehm}},
  \bibinfo{author}{\bibfnamefont{R.}~\bibnamefont{Durrer}}, \bibnamefont{and}
  \bibinfo{author}{\bibfnamefont{C.}~\bibnamefont{van~de Bruck}},
  \bibinfo{journal}{Phys. Rev.} \textbf{\bibinfo{volume}{D64}},
  \bibinfo{pages}{063504} (\bibinfo{year}{2001}), \eprint{hep-th/0102144}.

\bibitem[{\citenamefont{Sasaki et~al.}(2000)\citenamefont{Sasaki, Shiromizu,
  and Maeda}}]{Sasaki:1999mi}
\bibinfo{author}{\bibfnamefont{M.}~\bibnamefont{Sasaki}},
  \bibinfo{author}{\bibfnamefont{T.}~\bibnamefont{Shiromizu}},
  \bibnamefont{and} \bibinfo{author}{\bibfnamefont{K.-i.} \bibnamefont{Maeda}},
  \bibinfo{journal}{Phys. Rev.} \textbf{\bibinfo{volume}{D62}},
  \bibinfo{pages}{024008} (\bibinfo{year}{2000}), \eprint{hep-th/9912233}.

\bibitem[{\citenamefont{Langlois}(2000)}]{Langlois:2000ia}
\bibinfo{author}{\bibfnamefont{D.}~\bibnamefont{Langlois}},
  \bibinfo{journal}{Phys. Rev.} \textbf{\bibinfo{volume}{D62}},
  \bibinfo{pages}{126012} (\bibinfo{year}{2000}), \eprint{hep-th/0005025}.

\bibitem[{\citenamefont{van~de Bruck et~al.}(2000)\citenamefont{van~de Bruck,
  Dorca, Brandenberger, and Lukas}}]{vandeBruck:2000ju}
\bibinfo{author}{\bibfnamefont{C.}~\bibnamefont{van~de Bruck}},
  \bibinfo{author}{\bibfnamefont{M.}~\bibnamefont{Dorca}},
  \bibinfo{author}{\bibfnamefont{R.~H.} \bibnamefont{Brandenberger}},
  \bibnamefont{and} \bibinfo{author}{\bibfnamefont{A.}~\bibnamefont{Lukas}},
  \bibinfo{journal}{Phys. Rev.} \textbf{\bibinfo{volume}{D62}},
  \bibinfo{pages}{123515} (\bibinfo{year}{2000}),
  \eprint[http://arXiv.org/abs]{hep-th/0005032}.

\bibitem[{\citenamefont{Bridgman et~al.}(2002)\citenamefont{Bridgman, Malik,
  and Wands}}]{Bridgman:2001mc}
\bibinfo{author}{\bibfnamefont{H.~A.} \bibnamefont{Bridgman}},
  \bibinfo{author}{\bibfnamefont{K.~A.} \bibnamefont{Malik}}, \bibnamefont{and}
  \bibinfo{author}{\bibfnamefont{D.}~\bibnamefont{Wands}},
  \bibinfo{journal}{Phys. Rev.} \textbf{\bibinfo{volume}{D65}},
  \bibinfo{pages}{043502} (\bibinfo{year}{2002}),
  \eprint[http://arXiv.org/abs]{astro-ph/0107245}.

\bibitem[{\citenamefont{Riazuelo et~al.}(2002)\citenamefont{Riazuelo, Vernizzi,
  Steer, and Durrer}}]{Riazuelo:2002mi}
\bibinfo{author}{\bibfnamefont{A.}~\bibnamefont{Riazuelo}},
  \bibinfo{author}{\bibfnamefont{F.}~\bibnamefont{Vernizzi}},
  \bibinfo{author}{\bibfnamefont{D.}~\bibnamefont{Steer}}, \bibnamefont{and}
  \bibinfo{author}{\bibfnamefont{R.}~\bibnamefont{Durrer}}
  (\bibinfo{year}{2002}), \eprint{hep-th/0205220}.

\bibitem[{\citenamefont{Mukohyama}(2000{\natexlab{a}})}]{Mukohyama:2000ga}
\bibinfo{author}{\bibfnamefont{S.}~\bibnamefont{Mukohyama}},
  \bibinfo{journal}{Class. Quant. Grav.} \textbf{\bibinfo{volume}{17}},
  \bibinfo{pages}{4777} (\bibinfo{year}{2000}{\natexlab{a}}),
  \eprint{hep-th/0006146}.

\bibitem[{\citenamefont{Mukohyama}(2000{\natexlab{b}})}]{Mukohyama:2000ui}
\bibinfo{author}{\bibfnamefont{S.}~\bibnamefont{Mukohyama}},
  \bibinfo{journal}{Phys. Rev.} \textbf{\bibinfo{volume}{D62}},
  \bibinfo{pages}{084015} (\bibinfo{year}{2000}{\natexlab{b}}),
  \eprint{hep-th/0004067}.

\bibitem[{\citenamefont{Mukohyama}(2001)}]{Mukohyama:2001yp}
\bibinfo{author}{\bibfnamefont{S.}~\bibnamefont{Mukohyama}},
  \bibinfo{journal}{Phys. Rev.} \textbf{\bibinfo{volume}{D64}},
  \bibinfo{pages}{064006} (\bibinfo{year}{2001}), \eprint{hep-th/0104185}.

\bibitem[{\citenamefont{Deruelle et~al.}(2001)\citenamefont{Deruelle, Dolezel,
  and Katz}}]{Deruelle:2000yj}
\bibinfo{author}{\bibfnamefont{N.}~\bibnamefont{Deruelle}},
  \bibinfo{author}{\bibfnamefont{T.}~\bibnamefont{Dolezel}}, \bibnamefont{and}
  \bibinfo{author}{\bibfnamefont{J.}~\bibnamefont{Katz}},
  \bibinfo{journal}{Phys. Rev.} \textbf{\bibinfo{volume}{D63}},
  \bibinfo{pages}{083513} (\bibinfo{year}{2001}), \eprint{hep-th/0010215}.

\bibitem[{\citenamefont{Deffayet}(2002)}]{Deffayet:2002fn}
\bibinfo{author}{\bibfnamefont{C.}~\bibnamefont{Deffayet}},
  \bibinfo{journal}{Phys. Rev.} \textbf{\bibinfo{volume}{D66}},
  \bibinfo{pages}{103504} (\bibinfo{year}{2002}), \eprint{hep-th/0205084}.

\bibitem[{\citenamefont{Leong et~al.}(2002)\citenamefont{Leong, Challinor,
  Maartens, and Lasenby}}]{Leong:2002hs}
\bibinfo{author}{\bibfnamefont{B.}~\bibnamefont{Leong}},
  \bibinfo{author}{\bibfnamefont{A.}~\bibnamefont{Challinor}},
  \bibinfo{author}{\bibfnamefont{R.}~\bibnamefont{Maartens}}, \bibnamefont{and}
  \bibinfo{author}{\bibfnamefont{A.}~\bibnamefont{Lasenby}},
  \bibinfo{journal}{Phys. Rev.} \textbf{\bibinfo{volume}{D66}},
  \bibinfo{pages}{104010} (\bibinfo{year}{2002}), \eprint{astro-ph/0208015}.

\bibitem[{\citenamefont{Frolov and Kofman}(2002)}]{Frolov:2002va}
\bibinfo{author}{\bibfnamefont{A.}~\bibnamefont{Frolov}} \bibnamefont{and}
  \bibinfo{author}{\bibfnamefont{L.}~\bibnamefont{Kofman}}
  (\bibinfo{year}{2002}), \eprint{hep-th/0212327}.

\bibitem[{\citenamefont{Durrer and Kocian}(2003)}]{Durrer:2003rg}
\bibinfo{author}{\bibfnamefont{R.}~\bibnamefont{Durrer}} \bibnamefont{and}
  \bibinfo{author}{\bibfnamefont{P.}~\bibnamefont{Kocian}}
  (\bibinfo{year}{2003}), \eprint{hep-th/0305181}.

\bibitem[{\citenamefont{Carter}(1997)}]{Carter:1997pb}
\bibinfo{author}{\bibfnamefont{B.}~\bibnamefont{Carter}}
  (\bibinfo{year}{1997}), \eprint{hep-th/9705172}.

\bibitem[{\citenamefont{Ringeval}(2002)}]{Ringeval:2002qi}
\bibinfo{author}{\bibfnamefont{C.}~\bibnamefont{Ringeval}},
  \emph{\bibinfo{title}{Fermionic currents flowing along extended objects}}
  (\bibinfo{publisher}{Phd Thesis}, \bibinfo{address}{University Paris 6},
  \bibinfo{year}{2002}), \eprint{hep-ph/0211126}.

\bibitem[{\citenamefont{Carter and Peter}(1999)}]{Carter:1999hx}
\bibinfo{author}{\bibfnamefont{B.}~\bibnamefont{Carter}} \bibnamefont{and}
  \bibinfo{author}{\bibfnamefont{P.}~\bibnamefont{Peter}},
  \bibinfo{journal}{Phys. Lett.} \textbf{\bibinfo{volume}{B466}},
  \bibinfo{pages}{41} (\bibinfo{year}{1999}), \eprint{hep-th/9905025}.

\bibitem[{\citenamefont{Lanczos}(1924)}]{Lanczos:1924}
\bibinfo{author}{\bibfnamefont{K.}~\bibnamefont{Lanczos}},
  \bibinfo{journal}{Ann. Phys.} \textbf{\bibinfo{volume}{74}},
  \bibinfo{pages}{518} (\bibinfo{year}{1924}).

\bibitem[{\citenamefont{Sen}(1924)}]{Sen:1924}
\bibinfo{author}{\bibfnamefont{N.}~\bibnamefont{Sen}}, \bibinfo{journal}{Ann.
  Phys.} \textbf{\bibinfo{volume}{73}}, \bibinfo{pages}{365}
  (\bibinfo{year}{1924}).

\bibitem[{\citenamefont{Darmois}(1927)}]{Darmois:1927}
\bibinfo{author}{\bibfnamefont{G.}~\bibnamefont{Darmois}},
  \emph{\bibinfo{title}{M{\'e}morial des sciences math{\'e}matiques, fasicule
  25 chap. 5}} (\bibinfo{publisher}{Gauthier-Villars},
  \bibinfo{address}{Paris}, \bibinfo{year}{1927}).

\bibitem[{\citenamefont{Israel}(1966)}]{Israel:1966}
\bibinfo{author}{\bibfnamefont{W.}~\bibnamefont{Israel}},
  \bibinfo{journal}{Nuovo Cimento} \textbf{\bibinfo{volume}{48B}},
  \bibinfo{pages}{463} (\bibinfo{year}{1966}).

\bibitem[{\citenamefont{Misner et~al.}(1970)\citenamefont{Misner, Thorne, and
  Wheeler}}]{Misner:1970aa}
\bibinfo{author}{\bibfnamefont{C.~W.} \bibnamefont{Misner}},
  \bibinfo{author}{\bibfnamefont{K.~S.} \bibnamefont{Thorne}},
  \bibnamefont{and} \bibinfo{author}{\bibfnamefont{J.~A.}
  \bibnamefont{Wheeler}}, \emph{\bibinfo{title}{Gravitation}}
  (\bibinfo{publisher}{Freeman, W. H. and company}, \bibinfo{address}{San
  Francisco, US}, \bibinfo{year}{1970}).

\bibitem[{\citenamefont{Abramowitz and Stegun}(1970)}]{Abramovitz:1970aa}
\bibinfo{author}{\bibfnamefont{M.}~\bibnamefont{Abramowitz}} \bibnamefont{and}
  \bibinfo{author}{\bibfnamefont{I.~A.} \bibnamefont{Stegun}},
  \emph{\bibinfo{title}{Handbook of mathematical functions with formulas,
  graphs, and mathematical tables}} (\bibinfo{publisher}{National Bureau of
  Standards}, \bibinfo{address}{Washington, US}, \bibinfo{year}{1970}),
  \bibinfo{edition}{ninth} ed.

\bibitem[{\citenamefont{Durrer}(1994)}]{Durrer:1993db}
\bibinfo{author}{\bibfnamefont{R.}~\bibnamefont{Durrer}},
  \bibinfo{journal}{Fundamentals of Cosmic Physics}
  \textbf{\bibinfo{volume}{15,3}}, \bibinfo{pages}{209} (\bibinfo{year}{1994}),
  \eprint[http://arXiv.org/abs]{astro-ph/9311041}.

\bibitem[{\citenamefont{Stewart and Walker}(1984)}]{Stewart:1984}
\bibinfo{author}{\bibfnamefont{E.~D.} \bibnamefont{Stewart}} \bibnamefont{and}
  \bibinfo{author}{\bibfnamefont{M.}~\bibnamefont{Walker}}, in
  \emph{\bibinfo{booktitle}{Proceedings of the Royal Society}}
  (\bibinfo{publisher}{Royal Society}, \bibinfo{address}{London, UK},
  \bibinfo{year}{1984}), chap. \bibinfo{chapter}{A341}, p.~\bibinfo{pages}{49}.

\bibitem[{\citenamefont{Durrer et~al.}(2002)\citenamefont{Durrer, Kunz, and
  Melchiorri}}]{Durrer:2001cg}
\bibinfo{author}{\bibfnamefont{R.}~\bibnamefont{Durrer}},
  \bibinfo{author}{\bibfnamefont{M.}~\bibnamefont{Kunz}}, \bibnamefont{and}
  \bibinfo{author}{\bibfnamefont{A.}~\bibnamefont{Melchiorri}},
  \bibinfo{journal}{Phys. Rept.} \textbf{\bibinfo{volume}{364}},
  \bibinfo{pages}{1} (\bibinfo{year}{2002}), \eprint{astro-ph/0110348}.

\bibitem[{\citenamefont{Hu and White}(1997)}]{Hu:1997hp}
\bibinfo{author}{\bibfnamefont{W.}~\bibnamefont{Hu}} \bibnamefont{and}
  \bibinfo{author}{\bibfnamefont{M.~J.} \bibnamefont{White}},
  \bibinfo{journal}{Phys. Rev.} \textbf{\bibinfo{volume}{D56}},
  \bibinfo{pages}{596} (\bibinfo{year}{1997}), \eprint{astro-ph/9702170}.

\bibitem[{\citenamefont{Boehm}()}]{timon:2003}
\bibinfo{author}{\bibfnamefont{T.}~\bibnamefont{Boehm}},
  \emph{\bibinfo{title}{In preparation}}.

\bibitem[{\citenamefont{Gradshteyn and Ryzhik}(1965)}]{Gradshteyn:1965aa}
\bibinfo{author}{\bibfnamefont{I.~S.} \bibnamefont{Gradshteyn}}
  \bibnamefont{and} \bibinfo{author}{\bibfnamefont{I.~M.}
  \bibnamefont{Ryzhik}}, \emph{\bibinfo{title}{Table of Integrals, Series, and
  Products}} (\bibinfo{publisher}{Academic Press}, \bibinfo{address}{New York
  and London}, \bibinfo{year}{1965}).

\end{thebibliography}

\end{document}